\documentclass[twocolumn]{aastex62}

\usepackage{graphicx}
\usepackage{amsmath}
\usepackage{amssymb}
\usepackage{bm}
\usepackage{cleveref}
\usepackage{natbib}         
\usepackage{float}

\received{2019 March 27}
\revised{2019 June 7}
\accepted{2019 June 11}
\published{2019 August 1}

\shorttitle{{\it HST} Exospheric Mg II and Fe II in WASP-121\MakeLowercase{b}} 
\shortauthors{Sing et al.} 

\begin{document}

\title{The HST PanCET Program: Exospheric Mg II and Fe II in the Near-UV transmission spectrum of WASP-121b using Jitter Decorrelation}

\correspondingauthor{David K. Sing}
\email{dsing@jhu.edu}

\author[0000-0001-6050-7645]{David K. Sing}
\affil{Department of Earth \& Planetary Sciences, Johns Hopkins University, Baltimore, MD, USA}
\affil{Department of Physics \& Astronomy, Johns Hopkins University, Baltimore, MD, USA}

\author[0000-0002-5360-3660]{Panayotis Lavvas}
\affil{Groupe de Spectrom\'etrie Mol\'eculaire et Atmosph\'erique, Universit\'e de Reims, Champagne-Ardenne, CNRS UMR 7331, France}

\author[0000-0002-3891-7645]{Gilda E. Ballester}
\affil{Lunar and Planetary Laboratory, University of Arizona, Tucson, AZ 85721, USA}

\author[0000-0002-5637-5253]{Alain Lecavelier des Etangs}
\affil{Sorbonne Universit\'{e}s, UPMC Universit\'{e} Paris 6 and CNRS, UMR 7095, Institut d'Astrophysique de Paris, 98 bis boulevard Arago, F-75014 Paris, France}

\author[0000-0002-5251-2943]{Mark S.\ Marley}
\affil{NASA Ames Research Center, Moffett Field, California, USA}

\author[0000-0002-6500-3574]{Nikolay Nikolov}
\affil{Department of Earth \& Planetary Sciences, Johns Hopkins University, Baltimore, MD, USA}

\author[0000-0003-4047-2793]{Lotfi Ben-Jaffel}
\affil{Sorbonne Universit\'{e}s, UPMC Universit\'{e} Paris 6 and CNRS, UMR 7095, Institut d'Astrophysique de Paris, 98 bis boulevard Arago, F-75014 Paris, France}

\author[0000-0002-9148-034X]{Vincent Bourrier}
\affil{Observatoire de l'Universit\'{e} de Gen\`{e}ve, 51 chemim des Maillettes, 1290 Sauverny, Switzerland}

\author[0000-0003-1605-5666]{Lars A. Buchhave}
\affil{DTU Space, National Space Institute, Technical University of Denmark, Elektrovej 328, DK-2800 Kgs. Lyngby, Denmark}

\author[0000-0001-5727-4094]{Drake L. Deming}
\affil{Department of Astronomy, University of Maryland, College Park, Maryland, USA}

\author[0000-0001-9704-5405]{David Ehrenreich}
\affil{Observatoire de l'Universit\'{e} de Gen\`{e}ve, 51 chemim des Maillettes, 1290 Sauverny, Switzerland}

\author[0000-0001-5442-1300]{Thomas Mikal-Evans}
\affil{Kavli Institute for Astrophysics and Space Research, Massachusetts Institute of Technology, 77 Massachusetts Avenue, 37-241, Cambridge, MA 02139, USA}

\author[0000-0003-3759-9080]{Tiffany Kataria}
\affil{NASA Jet Propulsion Laboratory, 4800 Oak Grove Drive, Pasadena, CA 91109, USA}

\author[0000-0002-8507-1304]{Nikole K. Lewis}
\affil{Department of Astronomy and Carl Sagan Institute, Cornell University, 122 Sciences Drive, Ithaca, NY 14853, USA }

\author[0000-0003-3204-8183]{Mercedes L\'opez-Morales}
\affil{Center for Astrophysics | Harvard \& Smithsonian, 60 Garden Street, Cambridge, MA 02138, USA}

\author[0000-0003-1756-4825]{Antonio Garc\'ia Mu\~noz}
\affil{Zentrum f\"ur Astronomie und Astrophysik, Technische Universit\"at Berlin,
Berlin, Germany}

\author[0000-0003-4155-8513]{Gregory W. Henry}
\affil{Center of Excellence in Information Systems,  Tennessee State University, Nashville, TN  37209, USA}

\author[0000-0002-1600-7835]{Jorge Sanz-Forcada}
\affil{Centro de Astrobiolog\'{i}a (CSIC-INTA), E-28692 Villanueva de la Ca\~nada, Madrid, Spain}

\author[0000-0002-5547-3775]{Jessica J. Spake}
\affil{Physics and Astronomy, Stocker Road, University of Exeter, Exeter, EX4 3RF, UK}

\author[0000-0003-4328-3867]{Hannah R.\ Wakeford}
\affil{Space Telescope Science Institute, 3700 San Martin Drive, Baltimore, Maryland 21218, USA}

\collaboration{(the PanCET collaboration)}

\begin{abstract}
We present {\it HST} near-ultraviolet (NUV) transits of the hot Jupiter WASP-121b, acquired as part of the PanCET program.  Time series spectra during two transit events were used to measure the transmission spectra between 2280 and 3070 \AA\ at a resolution of 30,000.  Using {\it HST} data from 61 STIS visits, we show that data from HST's Pointing Control System can be used to decorrelate the instrument systematic errors (Jitter Decorrelation), which we used to fit the WASP-121b light curves.  The NUV spectrum show very strong absorption features, with the NUV white light curve found to be larger than the average optical and near-infrared value at 6-$\sigma$ confidence.  We identify and spectrally resolve absorption from the Mg\,{\sc ii} doublet in the planetary exosphere at a 5.9-$\sigma$ confidence level.  The Mg\,{\sc ii} doublet is observed to reach altitudes of $R_{pl}/R_{star}=0.284\pm0.037$ for the 2796 \AA\ line and $0.242\pm0.0431$ in the 2804 \AA\ line, which exceeds the Roche lobe size as viewed in transit geometry ($R_{\rm eqRL}/R_{star}$ = 0.158). We also detect and resolve strong features of the Fe\,{\sc ii} UV1 and UV2 multiplets, and observe the lines reaching altitudes of $R_{pl}/R_{star}\approx0.3$.  At these high altitudes, the atmospheric Mg\,{\sc ii} and Fe\,{\sc ii} gas is not gravitationally bound to the planet, and these ionized species may be hydrodynamically escaping or could be magnetically confined.  Refractory Mg and Fe atoms at high altitudes also indicates that these species are not trapped into condensate clouds at depth, which places constraints on the deep interior temperature.

\end{abstract}


\section{Introduction} \label{sec:intro}
Close-in exoplanets are exposed to immense stellar X-ray and extreme-UV (EUV) radiation. These photons ionize atmospheric species and deposit enormous heat through electron collisions, that can greatly expand the upper atmosphere and drive hydrodynamic outflow and mass escape \citep{2004Icar..170..167Y, 2007P&SS...55.1426G, 2009ApJ...693...23M, 2012MNRAS.425.2931O}.  
Ultraviolet observations of transiting exoplanets are sensitive to these uppermost microbar atmospheric layers, where atomic and ionized species can be detected.  
Thus-far, these detections include species such as H\,{\sc i} in Ly-$\alpha$, as well as O\,{\sc i}, C \,{\sc ii} in the Far Ultraviolet (FUV) and Mg\,{\sc i} in the near-ultraviolet (NUV, 2000 - 3000 \AA).  In the case of HD~209458b, $\sim$10\% UV transit depths were found, indicating an extended H\,{\sc i}, O\,{\sc i},  C \,{\sc ii}, and Mg\,{\sc i} exosphere \citep{2003Natur.422..143V, 2004ApJ...604L..69V, 2013A&A...560A..54V, 2010ApJ...717.1291L, 2015ApJ...804..116B, 2007ApJ...671L..61B, 2008ApJ...688.1352B}.   
Detections of H\,{\sc i} and O\,{\sc i} have also been made in HD 189733b 
\citep{2010A&A...514A..72L, 2012A&A...543L...4L, 2013A&A...553A..52B}. 
Some of the best examples of escaping exoplanet atmospheres occur in warm-Neptune mass planets.  For GJ~436b \citep{2015Natur.522..459E}, an extensive H\,{\sc i} atmosphere has been detected, including an extended cometary-like tail. The transit depth reaches $\sim$50\% at Ly-$\alpha$, showing H\,{\sc i} extending well beyond the Roche lobe and surviving photo-ionization \citep{2015Natur.522..459E, 2017A&A...605L...7L}.  Similarly, the warm Neptune GJ~3470b also shows an extended upper atmosphere of H\,{\sc i} \citep{2018A&A...620A.147B}, with transit absorption depths reaching 35\%.

The most highly irradiated hot Jupiters are expected to have vigorous atmospheric escape.  These exoplanets could be excellent probes of evaporation and photo-ionization, as metals such as Fe and Mg are not condensed into clouds \citep{2010ApJ...716.1060V, 2017MNRAS.464.4247W}, which would otherwise trap these atomic species in the lower atmosphere.  
Neutral and ionized Fe and Ti have been detected in the ultra-hot Jupiter KELT-9b \citep{2017Natur.546..514G} from high spectral resolution ground based observations \citep{2018Natur.560..453H} as have Mg\,{\sc i} and H\,{\sc i} in H$\alpha$ 
\citep{2018NatAs...2..714Y, 2019AJ....157...69C, 2019arXiv190502096H}. 
As ultra-hot Jupiters are rare and often located in systems far from the sun, the currently known population is inaccessible to FUV observations due to the prohibitively large absorption by the interstellar medium (ISM).  However, these very hot exoplanets are accessible in the NUV where there are numerous atomic spectral lines and the ISM does not pose such a large problem.
For the ultra-hot Jupiter WASP-12b, which has an equilibrium temperature of 2580 K, NUV spectrophotometry with the Cosmic Origins Spectrograph revealed strong broadband absorption signatures between 2539 and 2811 \AA.  These signatures have been interpreted as a continuum of metal lines absorbing the entire NUV region \citep{2010ApJ...714L.222F, 2013ApJ...766L..20F, 2012ApJ...760...79H}, and excess transit depths were seen at Mg\,{\sc ii} and Fe\,{\sc ii} wavelengths.  Possible early ingress signatures in H$\alpha$ and Na have also been seen at optical wavelengths \citep{2018AJ....156..154J}.
An early ingress was also found, which could be a signature of material overflowing the Roche lobe at the L1 Lagrangian point or a magnetospheric bow shock 
\citep{2010ApJ...721..923L, 2010ApJ...722L.168V, 2011MNRAS.416L..41L, 2013ApJ...764...19B}.  
However, Mg\,{\sc ii} and Fe\,{\sc ii} do not exist in the stellar wind or corona, which poses problems for the bow-shock interpretation \citep{2014ApJ...785L..30B}.  
These observations were followed up by additional WASP-12b COS observations \citep{2015ApJ...803....9N}, which also showed significant NUV absorption but did not find evidence for the previously claimed early ingress (see also \citealt{2016MNRAS.459..789T}).

WASP-121b is an ultra-hot Jupiter discovered by \cite{2016MNRAS.458.4025D}, with a dayside equilibrium temperature above 2400 K.  The planet has a mass of 1.18$\pm$0.06 M$_{\rm J}$, a large inflated radius of $\sim$1.7 R$_{\rm J}$, and is in a short orbital period around a bright (V$_{mag}$=10.5) F6V star.  WASP-121b is extremely favourable to atmospheric measurements.  The planet's dayside emission spectra has been measured in the near-IR with {\it HST} WFC3, where a stratosphere has been found exhibiting spectroscopically resolved emission features of H$_2$O \citep{2017Natur.548...58E}.  The transmission spectra acquired with {\it HST} STIS and WFC3 revealed molecular absorption due to H$_2$O and likely VO \citep{2016ApJ...822L...4E, 2018AJ....156..283E}.   In addition, strong NUV absorption signatures between 3000 and 4500 \AA\ were found in the STIS G430L transmission data, which \cite{2018AJ....156..283E} interpreted as a possible SH signature or some other absorber.  NUV observations of WASP-121b have also been made with the SWIFT telescope \citep{2019arXiv190110223S}, where a tentative excess absorption signature has also been seen.

Here we present Hubble Space Telescope ({\it HST}) transmission spectrum results for WASP-121b from the Panchromatic Exoplanet Treasury (PanCET) program ({\it HST} GO-14767 P.I.s Sing \& L\'opez-Morales), which consists of multi-wavelength transit and eclipse observations for 20 exoplanets.  
In this work, we present new {\it HST} NUV transit observations with the Space Telescope Imaging Spectrograph (STIS).
We describe a new method to decorrelate time series HST light curves in Section \ref {sec:jitter},
present our observations in Section 3, give the analysis of the
transit light curves in Section 4, discuss the results in Section 5 and conclude in Section 6.

\section{Hubble Jitter Decorrelation} \label{sec:jitter}
{\it HST} spectrophometric light curves taken with the Space Telescope Imaging Spectrograph (STIS) show instrument-related systematic effects that are widely thought to be caused by the thermal breathing of {\it HST}.  
The thermal breathing trends cause the point-spread function (PSF) to change repeatedly for each 90 minute spacecraft orbit around the Earth, producing corresponding photometric changes in the light curve \citep{2001ApJ...552..699B}.  These systematic trends have widely been removed by 
a parameterized deterministic model, where the photometric trends are found to correlate with a number
$n$ of external detrending parameters (or optical state parameters, $\mathbf{x}$).
These parameters describe changes in the instrument or other external
factors as a function of time during the observations, and are fit
with a coefficient for each optical state parameter, $p_n$, to model and detrend the photometric
light curves.  In the case of {\it HST}  STIS data, external detrending parameters including the 96
minute {\it HST}  orbital phase, $\phi_{HST}$, the $X_{psf}$ and $Y_{psf}$ detector
position of the PSF, and the wavelength shift $S_{\lambda}$ of the
spectra have been identified as optical state parameters (\citealt{2001ApJ...552..699B,2011MNRAS.416.1443S}).  This set of detrending parameters (hereafter called the `traditional model') have been widely used to fit {\it HST} transit and eclipse light curves (e.g. \citealt{2013MNRAS.434.3252H,  2013MNRAS.435.3481W, 2014MNRAS.437...46N, 2015MNRAS.450.2043D, 2016Natur.529...59S, 2018AJ....155...66L}), and even non-parametric models like Gaussian Processes still rely on these detrending parameters of the traditional model as inputs \citep{2017MNRAS.467.4591G, 2017ApJ...847L...2B}.  In addition to the STIS CCD with the G430L, G750L and G750M STIS gratings, these detrending parameters have been used to fit STIS Echelle E230M data, that uses the Multi-Anode Microchannel Array (NUV-MAMA) detector.

A potential source of astrophysical noise from our main-sequence FGK target stars is expected to be photometric variations due to stellar activity, with star spots modulating the total brightness of the star in a quasi-periodic fashion on timescales similar to the rotation period of the star (typically ranging from a few days to a few weeks).  However, our {\it HST}  transit observations only span 6.4 to 8 hours (four or five {\it HST} orbits), which is short compared to the stellar rotational period.  The small fraction of the stellar rotation period observed strongly limits the photometric amplitude of any rotational modulation in our transit light curves, and makes any quasi-periodic variations appear as linear changes in the baseline stellar flux.  We estimate that highly active stars variable at the 1\% level in the optical over 10 day periods (similar to HD189733A) would be expected to have photometric stellar activity change the baseline stellar flux during the {\it HST} visit by $\sim$600 ppm or less.  Inactive stars such as WASP-12A, which is photometrically quiet below 0.19\% \citep{2013MNRAS.436.2956S}, would be expected to have an upper limit to a stellar activity signal of 100 ppm during our transit observations, which is below the photon noise limit of a typical STIS CCD white-light-curve exposure.  These potential astrophysical noise sources can be modeled by fitting a linear function of time to the transit light curves, $\phi_{t}$.

There are still limitations with the traditional model in detrending instrument-related systematic effects.  Notably, the first orbit in {\it HST} STIS visits has generally been discarded as it takes one spacecraft orbit for the telescope to thermally relax, which compromises the photometric stability of the first orbit of each {\it HST} visit beyond the limits of the traditional model.  
 To help reduce the level of systematic errors in {\it HST} STIS measurements, in this paper we investigated additional optical state parameters which could be used to help decorrelate transit or eclipse light curves.  Rather than attributing all the trends to a thermal breathing effect, we explored parameters under the assumption that the stability of the telescope pointing itself (and subsequent effects such as slit light losses) is a limiting factor.  
The jitter files are products of the Engineering Data Processing System (EDPS), which describes the performance of  {\it HST}'s Pointing Control System (PCS) during the duration of an observation.\footnote{See http://www.stsci.edu/hst/observatory/pointing/ for additional details.}  The files contain a determination of the telescope pointing during an observation utilizing the Fine Guidance Sensor (FGS) guidestar measurements to map the telescope's focal plane onto a RA and DEC.  The determination is accurate to the level of the guidestar positional uncertainty and the knowledge of the telescope's focal plane geometry, which is typically at the sub-pixel level.  The data contained in the jitter files are time-tagged and recorded in 3 second averages, thus providing suitable fidelity to describe the pointing during a transiting exoplanet observations, which have science frame exposure times typically of several minutes.

For an {\it HST} visit, we extracted the time-tagged engineering information from the jitter files (extension {\it\_jit.fits}) for each science exposure by taking the median value of each measured quantity between the beginning and end date of the exposure.  We found using the median made the values less sensitive to occasional bad values recorded in the jitter files (a known issue).  For the jitter file measurement of the {\it HST} sub-point longitude, measured in degrees between 0 and 360, care was taken to prevent a value discontinuity in the time-series at 360 degrees.  For use in decorrelation as optical state vectors, we then normalized each measured quantity by removing the mean value from each variable and then dividing the values by the standard deviation (see Appendix \ref{sec:W12appendix} for additional details).

\subsection{Global STIS Light Curve-Jitter Correlation Trends}\label{sec:trends}
The jitter files include 28 different engineering measurements (see Table \ref{tab:tablejit}).  To determine the suitability of each of the jitter optical state vectors for decorrelating transit or eclipse light curves, we measured the linear Pearson correlation coefficient, $R$, between each jitter vector and the out-of-transit white-light-curve photometric data for 61 separate {\it HST} visits covering close to 300 spacecraft orbits.  Specifically, we used all of the STIS G430L and G750L data from 23 visits of program GO-12473 (P.I. Sing), G430L and G750L data from 34 visits of GO-14767 (P.I.s Sing \& L\'opez-Morales), and four G430L visits from GO-14797 (P.I. Crossfield).  The data from each {\it HST} STIS visit was reduced in the same manner as detailed in \cite{2016Natur.529...59S}, and we summed the counts of each extracted spectra to produce a white-light-curve for each visit.  Using the out-of-transit data, the percent of variance in common between jitter decorrelation parameters and the STIS white-light-curve photometry, $R^2$, was determined for each of the 61 visits.  We also measured the quantity for the traditional detrending variables ($\phi_{HST}$, $X_{psf}$, $Y_{psf}$ and  $S_{\lambda}$) with the results given in Table \ref{tab:tablejit}.  As a point of comparison, the individual $R^2$ values for a G430L eclipse of WASP-12b are reported in addition to the two STIS E230M transits of WASP-121b which are visits 97 and 98 of the PanCET program.  The raw white-light curve of the STIS G430L eclipse of WASP-12b reaches a photometric precision of 979 ppm, as measured by the standard deviation of the light curve about the mean, which is a factor of 6.4$\times$ higher than expected from the photon noise levels.

An additional systematic trend is seen in the STIS CCD time-series data, where the first exposure of each spacecraft orbit is consistently found to exhibit significantly lower fluxes than the remaining exposures (\citealt{2001ApJ...552..699B,2011MNRAS.416.1443S}).  Attempts were made to mitigate this effect in the observational setup by employing a quick 1 second integration \citep{2015MNRAS.446.2428S}, which is discarded.  However, higher overall correlation values were found between the STIS CCD white light curves and the optical state vectors if, in addition, the first long science exposure of every orbit was also systematically discarded from the analysis.  We suspect the 1 second integration technique is not an overall effective method to mitigate the first-orbit-exposure systematic, and the first long science exposure of any STIS CCD orbit in a time series analysis should be treated with caution.  For the correlation analysis and for all subsequent CCD light curve fitting throughout this paper, we chose to discard the first long science exposure of each {\it HST} orbit.

From the correlation analysis, we identify several pairs of jitter vectors that are often found to correlate well with the STIS data ($R^2\gtrsim30\%$, also see Appendix).  The roll of the telescope along the V2 and V3 axis ($V2\_roll$, $V3\_roll$) and the right ascension and declination of the aperture reference $RA$ and $DEC$ are two pairs of vectors that are both related to the pointing of the telescope and in some datasets are found to highly correlate with the data with $R^2$ values as high as 90\% seen in some datasets.  In addition, the {\it HST} sub-point latitude and longitude ($Lat$ and $Long$ respectively) are also seen to have large correlations.  The latitude and longitude are similar in nature (but not identical) to $\phi_{HST}$, which is already included in the traditional model.  From {\it HST} visit to visit, the correlation values themselves for any particular vector are found to change by large amounts, which is reflected in Table \ref{tab:tablejit} by the large standard deviation values, $\sigma_{R^2}$.  However, in retrospect a large spread in correlation values related to the telescope's position is to be expected.  Different targets across the sky observed at different telescope orientation values will not necessarily favour any particular telescope axis or $RA$ and $DEC$ direction, but rather the actual vector direction(s) the target PSF takes with respect to the instrument during a transit observation.  We note a general trend where the $R^2$ values in visits with relatively good pointing show lower levels of correlation (e.g. visit 98), while visits with poorer quality pointing show higher levels of correlation (e.g. visit 97).

\startlongtable
\begin{deluxetable}{c|ccc|cc}
\tablecaption{Jitter engineering data optical state vectors (column 1) and the correlation $R^2$ values (in \%) with the white light curve photometry for the visits covering a STIS/G430L eclipse of WASP-12 (column 1) and the STIS/E230M data of WASP-121 for visits 97 and visit 98 (columns 3 and 4 respectively).  The average $R^2$ value from 61 {\it HST} STIS G430L \& G750L CCD visits, $\overline{R^2}$, is also reported in column 5, and the standard deviation of $R^2$ in column 6.  \label{tab:tablejit}}
\tablehead{
\colhead{Vector} &\colhead{W-12} &\colhead{W-121} &\colhead{W-121} &$\overline{R^2}$ &\colhead{$\sigma_{\overline{R^2}}$} 
}
\colnumbers
\startdata
$\phi_{HST}$ & 36 & 54& 17 &34 &28 \\
$S_{\lambda}$& 86 &0 &2 &30 &27 \\
$X_{psf}$ &26 &30 & 1 &19 &22 \\
$Y_{psf}$ &31 &30 & 1 &16 &20 \\
$V2\_dom$& 16 & 13& 6& 7 & 9 \\
$V3\_dom$& 1 & 1 & 0 & 6 & 8 \\
$\boldsymbol{V2\_roll}$ &57 & 36 &7 &36 &29 \\
$\boldsymbol{V3\_roll}$ &54 & 34 &9 &35 &29 \\
$SI\_V2\_AVE$& 0 & 11 &19  & 4 & 5\\
$SI\_V2\_RMS$ &56 & 8 &0 &19 &20 \\
$SI\_V2\_P2P$ &58 & 15 &0 &19 &20 \\
$SI\_V3\_AVE$ &4 & 14 & 15 & 5 & 6 \\
$SI\_V3\_RMS$ &55 & 29 &6 &12 &15 \\
$SI\_V3\_P2P$ &55 &5 & 2&13 &15 \\
$\boldsymbol{RA}$ &55 & 37&8 &31 &28 \\
$\boldsymbol{DEC}$ &57 & 36 &7 &29 &27 \\
$Roll$ &56 & 50 &13 &14 &16 \\
$LimbAng$ &8 & 0 &3  &10 &14 \\
$TermAng$ &0 & 7& 3 & 8 & 9 \\
$LOS\_Zenith$ &8 & 0 &3 &10 &14 \\
$\boldsymbol{Lat}$ &13 &41 &7  &20 &21 \\
$\boldsymbol{Long}$ & 76 & 6 &0 &38 &29 \\
$Mag\_V1$& 1 & 22& 8 &13 &15 \\
$Mag\_V2$& 12 & 4& 0 &16 &19 \\
$Mag\_V3$& 12 &29 & 4 &15 &16 \\
\enddata
\end{deluxetable}

Trends with telescope position are consistent with photometric systematics caused by slit light losses.  If correct, the trends appear even when using slit sizes that are much greater than the size of the PSF.  Our hypothesis is that the known telescope breathing leads to small changes in the position of the PSF on the detector (usually at the sub-pixel level), which largely manifest themselves as slit light losses.  Even if the central PSF is well centered on a much wider slit, diffraction spikes and the wide wings of the PSF could contribute to light losses.  If the dominant source of systematics with STIS are position related slit light losses, assuming similar guiding performance, then presumably there could be larger systematics when using small slits such as the 0.2\arcsec$\times$0.2\arcsec\ used with the STIS E230M compared to the 52\arcsec$\times2$\arcsec\ slit used with the CCD.  In addition, there would be larger systematics in visits with poorer guiding performance, which appears to be the case.  Several STIS visits in the PanCET program had guide star acquisition problems, which resulted in reduced pointing accuracy and light curves with comparably larger systematic trends.
We also observe that the jitter vectors ($V2\_roll$, $V3\_roll$) and ($RA$ and $DEC$) often have the same general trends as the traditional vectors ($S_{\lambda}$, $X_{psf}$, and $Y_{psf}$), which are measured directly from the spectra.  Similar trends are generally expected, as changes in the PSF position (measured through the telescope position in the jitter files) will also be recorded by changes in the placement of the target stellar spectra on the detector.  These vectors are not identical, however, as the detector-measured vectors will also be sensitive to further detector effects such as the pixel-to-pixel flat fielding, while the jitter vectors are able to record the telescope position regardless of whether the target PSF is contained within the slit. 

As a test, we implemented Jitter detrending on the STIS G430L eclipse data of WASP-12b, finding good agreement in the eclipse depth with \cite{2017ApJ...847L...2B} who used a Gaussian process method to decorrelate the time series data (see Appendix \ref{sec:W12appendix}). 
However, several improvements can be seen utilizing Jitter Detrending.  The best-fit measured eclipse depth was found to be positive while \cite{2017ApJ...847L...2B} found a negative value.
In addition, with Jitter Detrending the first orbit was successfully recovered for use in the analysis, while \cite{2017ApJ...847L...2B} had to discard the orbit.

\section{Observations} \label{sec:obs}

\subsection{Hubble Space Telescope STIS NUV spectroscopy}\label{sec:eclipseobs}
\begin{figure*}
	\centering
	\includegraphics[width=\linewidth]{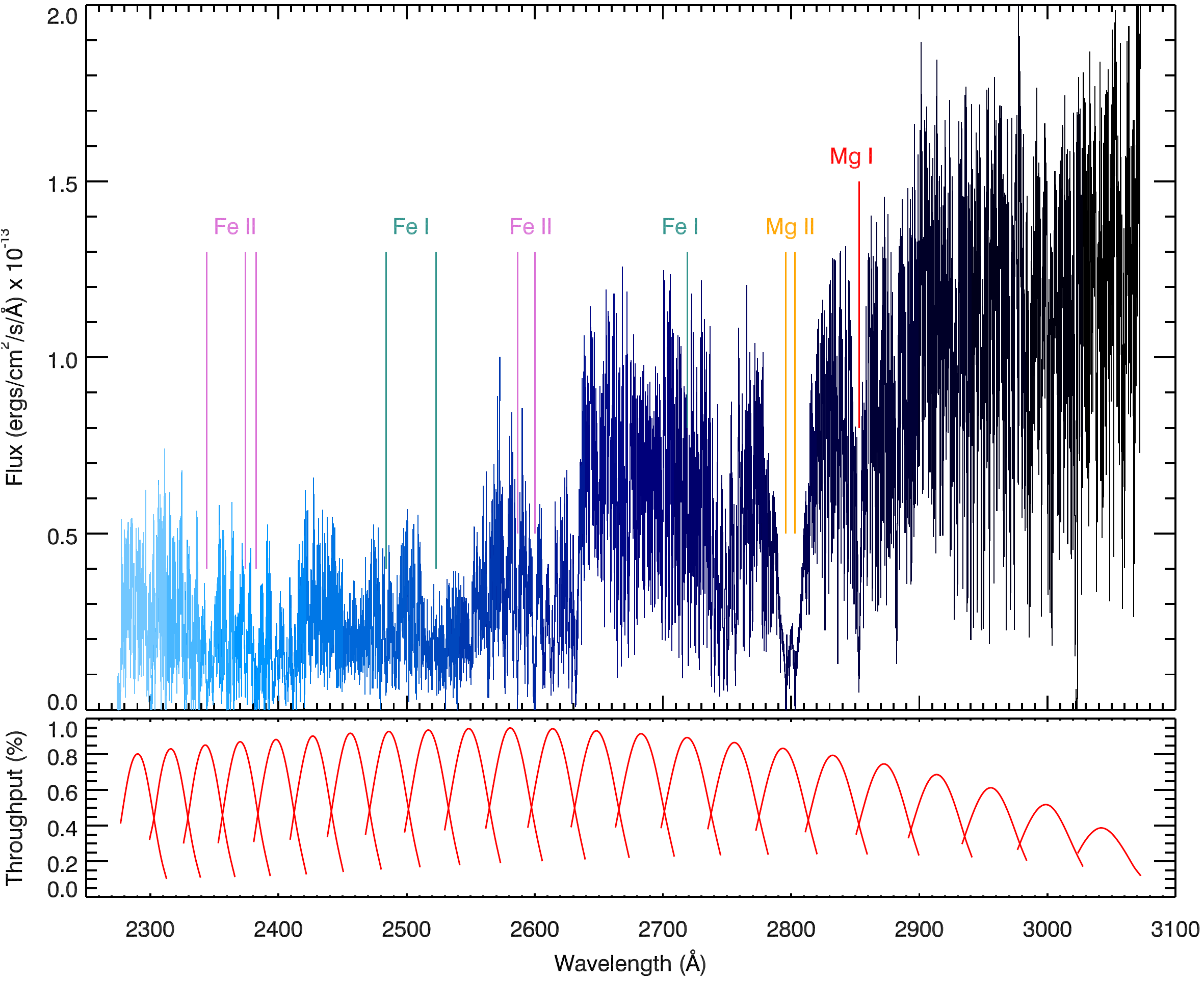}
	\caption{(Top) Flux calibrated out-of-transit spectrum of WASP-121A, each order is plotted in a different color.  Several resonant lines of Mg and Fe are indicated.  (Bottom) The instrumental throughput of each order. }\label{fig:specW121A}
\end{figure*}

We observed two transits of WASP-121b with the {\it HST} STIS E230M echelle grating during 23 February 2017 (visit 97) and 10 April 2017 (visit 98).  The observations were conducted with the NUV-Multi-Anode Microchannel Array (NUV-MAMA) detector using the Echelle E230M STIS grating and a square 0.2\arcsec$\times$0.2\arcsec\ entrance aperture.  The E230M spectra has resolving power of $\lambda$/($2\Delta$$\lambda$) = 30,000 and was configured to the 2707 \AA\ setting to cover the wavelength ranges between 2280 and 3070 \AA\ in 23 orders (see Fig. \ref{fig:specW121A}).  The resulting spectra have a dispersion on the detector of approximately 0.049 \AA/pix.  Both transits covered five {\it HST} spacecraft orbits, with the transit event occurring in the third and fourth orbits.  During each {\it HST}  orbit, the spectra were obtained in TIME-TAG mode, where the position and detection time of every photon is recorded in an event list, which has 125 microsecond precisions.

We sub-divided the spectrum of each {\it HST} spacecraft orbit into sub-exposures with the {\small Pyraf} task {\small inttag}, each with a duration of 274.01996 seconds.  This allowed orbits 2 through 5 to be divided into 10 sub-exposures each, while the first orbit was divided into 7 sub-exposures.  These sub-exposures were then each reduced with {\small CALSTIS} version 3.17, which includes the calibration steps of localization of the orders, optimal order extraction, wavelength calibration, and flat field corrections.
The mid-time of each exposure was converted into BJD$_{TDB}$ for use in the transit light curves \citep{2010PASP..122..935E}.  

\section{Analysis}
\subsection{STIS E230M light curve fits}\label{sec:lcfits}

The light curves were modeled with the analytical transit models of \cite{2002ApJ...580L.171M}. 
For the white-light curves, the central transit time, planet-to-star radius ratio, stellar baseline flux, and instrument systematic trends were fit simultaneously, with flat priors assumed.  The inclination and stellar density (or equivalently the semi-major axis to stellar radius ratio, $a/R_{star}$) were held fixed to the values found in \cite{2018AJ....156..283E}, as fitting for these parameters found values consistent with the literature, though with an order of magnitude lower precision.  For example, we find a value of $a/R_{star}$=3.63$\pm$0.16 from the NUV white light curve of visit 98, while \cite{2018AJ....156..283E} reports a value of 3.86$\pm$0.02.  Compared to the NUV data here, the study of \cite{2018AJ....156..283E} benefits from much higher photometric precision in a larger dataset with a wider array of wavelengths, including STIS data with complete phase coverage of the transit, which constrains the planet's orbital system parameters to a much greater degree than the NUV data alone.

\restylefloat*{figure}
\begin{figure}[ht]
\begin{minipage}[b]{.47\textwidth}
\centering
\includegraphics[width=\linewidth]{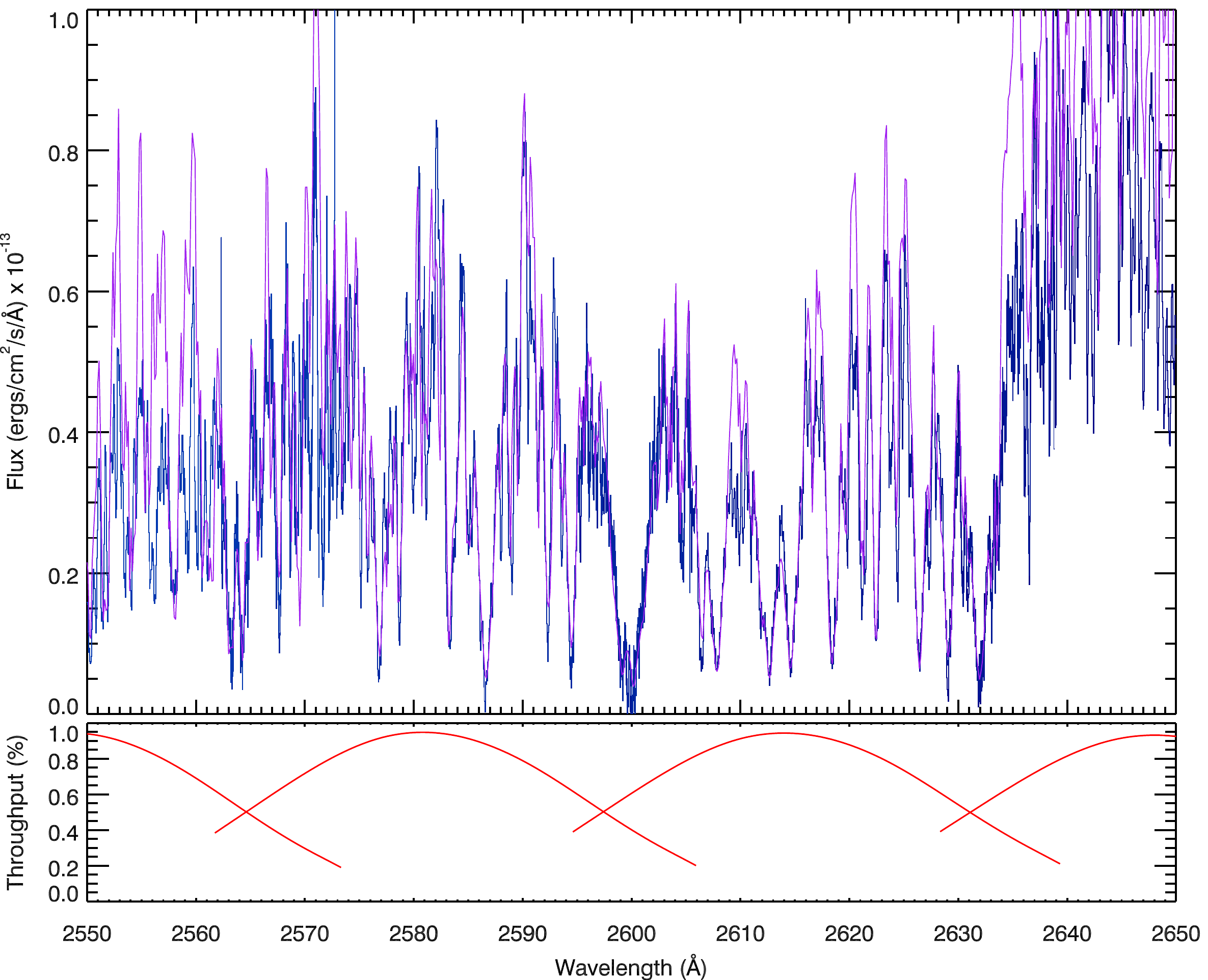}
\end{minipage}
\hfill
\begin{minipage}[b]{.47\textwidth}
\centering
\includegraphics[width=\linewidth]{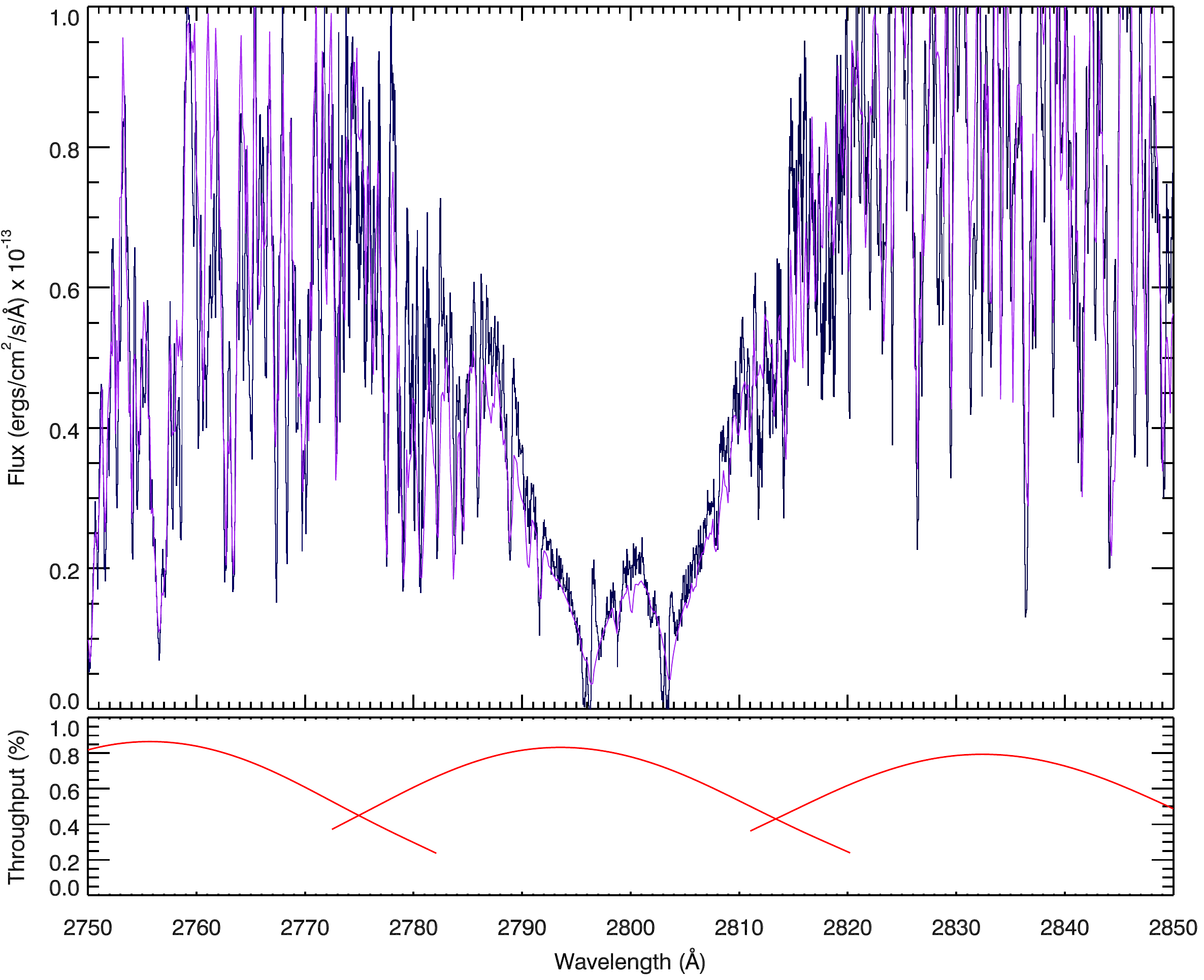}
\vspace*{-0.5\baselineskip}
\caption{Same as \ref{fig:specW121A}, but zoomed in on two wavelength regions covering a Fe\,{\sc ii} line at 2600 \AA\ and the Mg\,{\sc ii} doublet at 2796.35 and 2803.53 \AA\ with the 3D Stagger-grid model overplotted (purple).}\label{fig:3Dmodel}
\end{minipage}
\end{figure}

The total parameterized 
model of the flux measurements over time, $f(t)$, was 
modelled as a combination of the theoretical transit model, $T(t, \mathbf{\theta})$
(which depends upon the transit parameters $\mathbf{\theta}$), the total
baseline flux detected from the star, $F_0$, and the instrument systematics 
model $S(\mathbf{x})$ giving,
\begin{equation} 
f(t)=T(t, \theta)\times F_0 \times S(\mathbf{x}).
\end{equation} 
Based on the results of section \ref{sec:jitter} and Appendix \ref{sec:W12appendix}, for our most complex systematics error model tested, we included a linear baseline time trend, $\phi_{t}$, as well as the traditional optical state vectors ($\phi_{t}$, $\phi_{HST}$, $\phi_{HST}^2$, $\phi_{HST}^3$, $\phi_{HST}^4$, $X_{psf}$, $Y_{psf}$ \&  $S_{\lambda}$) and jitter vectors ($V2\_roll$, $V3\_roll$, $RA$, $DEC$, $Lat$, \& $Long$) resulting in up to fourteen total terms used to describe $S(\mathbf{x})$ depending on the dataset in question as described below.

The errors on each datapoint were initially set to the pipeline values, which
is dominated by photon noise.
The best-fitting parameters were determined 
simultaneously with a Levenberg-Marquardt (L-M) least-squares algorithm 
\citep{2009ASPC..411..251M} 
using the unbinned data.
After the initial fits, the uncertainties for each data point were
rescaled to match the standard deviation of the residuals.  A further scaling was also applied to account for any measured systematic errors correlated in time (`red noise').  After rescaling the error-bars, the light-curves were then refit, thus taking into account any underestimated
errors in the data points.  

The red noise was measured by checking whether the binned residuals followed a $N^{-1/2}$ relation, when
binning in time by $N$ points.  In the presence of red noise, the
variance can be modelled to follow a $\sigma^2=\sigma_{\rm
  w}^2/N+\sigma_{\rm r}^2$ relation,
where $\sigma_{\rm w}$ is the uncorrelated white noise component, and
$\sigma_{\rm r}$ characterizes the red noise \citep{
2006MNRAS.373..231P}.  For our best-fitting models for $S(\mathbf{x})$, we did not find evidence for substantial rednoise.

The uncertainties on the fitted parameters were calculated using the
covariance matrix from the Levenberg-Marquardt algorithm, which 
assumes that the probability space around the best-fit solution is
well-described by a multivariate Gaussian distribution.  Previous
analyses of other {\it HST} transit observations 
\citep{2012ApJ...747...35B, 2013ApJ...778..183L, 2013MNRAS.436.2956S,
  2014MNRAS.437...46N}  
have found this to be a good approximation when fitting {\it HST} STIS data.  We 
also computed uncertainties with a Markov Chain Monte Carlo analysis
\citep{2013PASP..125...83E}, 
which does not assume any functional form for
this probability distribution.  In each case, we found equivalent
results between the MCMC and the Levenberg-Marquardt algorithm for
both the fitted parameters and their uncertainties, as the posterior distributions were found to be Gaussian.  Inspection of the
2D probability distributions from both methods indicate that there are
no significant correlations between the systematic trend parameters
and the planet-to-star radius contrast.    

\subsection{Limb darkening}\label{sec:Limbd}

The effects of stellar limb-darkening are strong at NUV wavelengths. To account for the effects of limb-darkening on the NUV transit light
curve, we adopted the four parameter non-linear limb-darkening law, 
calculating the coefficients as described in \cite{2010A&A...510A..21S}. 
For the model, we used the 3D stellar model from the Stagger-grid \citep{2015A&A...573A..90M} with the model ($T_{eff}$=6500, log~$g$=4, [Fe/H]= 0.0), which was closest to the measured values of WASP-121A in effective temperature, gravity, and metallicity  ($T_{eff}$= 6460$\pm$140 K, log10 g = 4.242$\pm$0.2 cgs, [Fe/H] = +0.13$\pm$0.09 dex; \citealt{2016MNRAS.458.4025D}).  
The Stagger-grid 3D models are calculated at a resolution of $R=$20,000 which is close to the native resolution of the E230M data ($R=$30,000), and contains a fully line-blanketed NUV region that matches the flux-calibrated data of WASP-121A well (see Fig. \ref{fig:3Dmodel}).  
We included the individual responses from each order in the echelle spectra and converted the stellar model spectral wavelengths from air to vacuum.
We also applied a wavelength shift to the data to take into account the systemic radial velocity of the system, 38.36 km/s \citep{2018A&A...616A...1G}, shifting the star to the rest frame.

In a test to see how well the 3D models performed, we fit visit 98 with a 3-parameter limb-darkening law (see \citealt{2010A&A...510A..21S}) and let the linear coefficient, $c_2$, fit freely while the other two non-linear parameters were fixed to the model values.  We found good agreement at the 1-$\sigma$ level between the fit coefficient $c_2$=0.881$\pm$0.215 and the theoretical value of 1.077.  
In addition, the fit $R_{pl}/R_{star}=0.1415\pm0.0050$ was also consistent (within 1-$\sigma$) when fitting for $c_2$ vs fixing the limb-darkening (see Section \ref{sec:NUVwlc}).
For the remainder of the study, we fixed the limb-darkening coefficients to the model values.  We note
that using the second-order Akaike Information Criterion (AICc) for model selection (see Appendix \ref{sec:W12appendix}), that the AICc favoured fixing the limb-darkening parameters to the model values.

\subsection{Updated ephemeris with NASA TESS data}\label{sec:tess}
\begin{figure}
\begin{centering}
\includegraphics[width=0.49\textwidth]{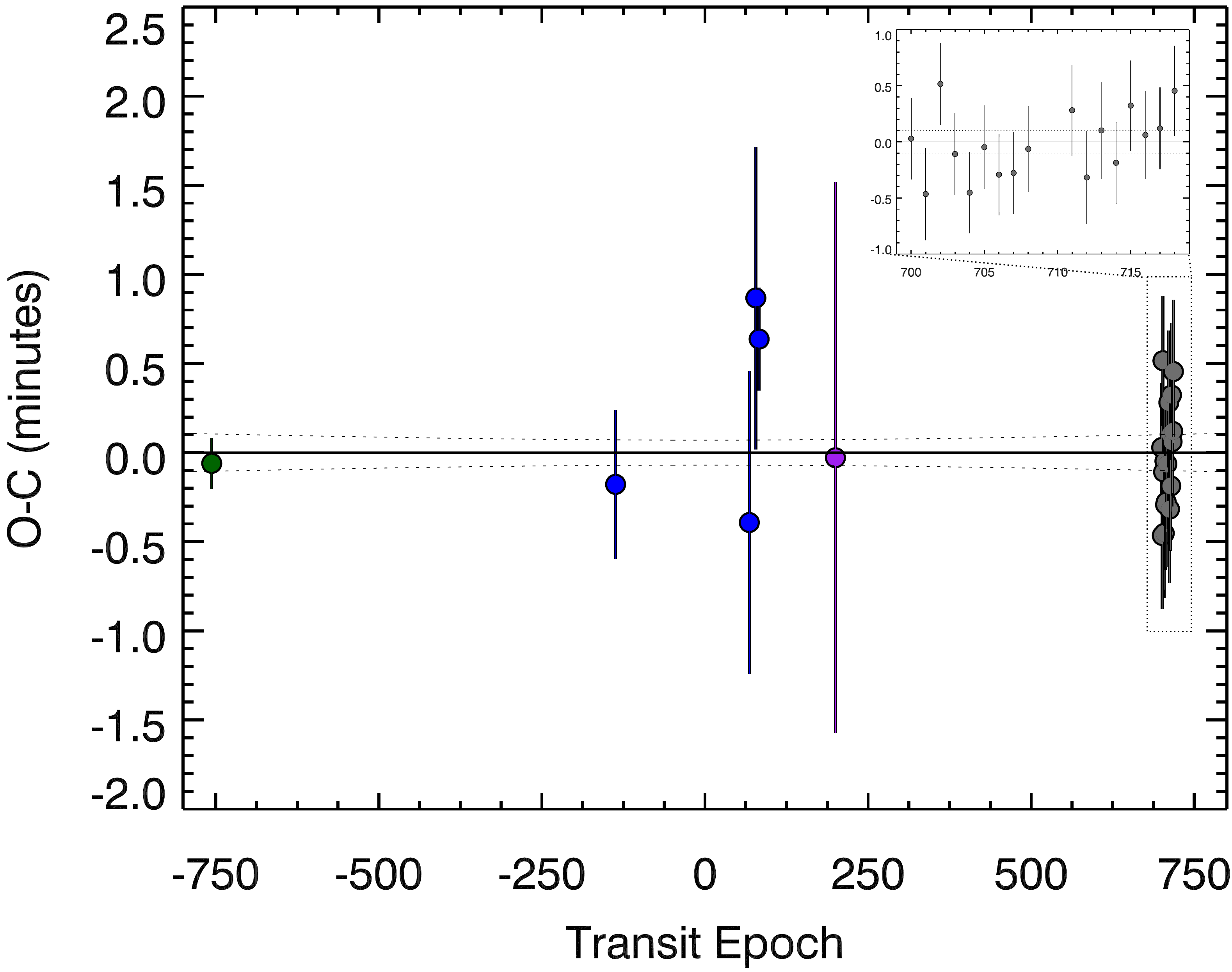}
\par\end{centering}
	\caption{Observed - Calculated mid-transit times of WASP-121b.  Included is the results from \cite{2016MNRAS.458.4025D} (dark green) and \cite{2018AJ....156..283E} (blue) as well as the the NUV transit time (purple) and TESS times (grey) from this work.  A zoom-in around the TESS mid-transit times is also shown. The 1-$\sigma$ error envelope on the ephemeris is plotted as the horizontal dashed lines.} \label{fig:timing}
\end{figure}

The Transiting Exoplanet Survey Satellite (TESS) observed WASP-121b in camera 3 from 7 Jan 2019 to 2 Feb 2019, and we obtained the timeseries photometry though the Mikulski Archive for Space Telescopes' exo.MAST\footnote{https://exo.mast.stsci.edu} web-service.  We fit the TESS WASP-121b transits using the Presearch Data Conditioning (PDC) light curve, which has been corrected for effects such as non-astrophysical variability and crowding \citep{2016SPIE.9913E..3EJ, 2012PASP..124..985S}.
From the timeseries, we removed all of the points which were flagged with anomalies.  The timeseries Barycentric TESS Julian Dates (BTJD) were converted to BJD$_{\rm TDB}$ by adding 2,457,000 days. The TESS light curve contains data for 16 complete transits of WASP-121b and one partial transit.  For each transit in the light curve, we extracted a 0.5 day window centered around the transits and fit each transit event individually.
We fit the data using the model as described in Sections \ref{sec:lcfits} and \ref{sec:Limbd}, though only included a linear baseline time trend, $\phi_{t}$, for $S(\mathbf{x})$.  We find a weighted-average value of $R_{pl}(\rm{TESS})/R_{star}=0.12342\pm0.00015$, which is in good agreement with the {\it HST} transmission spectrum of \cite{2018AJ....156..283E}.  We converted the transit times of \cite{2018AJ....156..283E} and \citep{2016MNRAS.458.4025D} to BJD$_{\rm TDB}$ using the tools from \citep{2010PASP..122..935E} and fit these along with the TESS transit times and visit 98 from Section \ref{sec:NUVwlc} with a linear function of the period $P$ and transit epoch $E$,
\begin{equation}
T(E)=T_0+EP.
\end{equation}
We find a period of  $P$=1.2749247646 $\pm$ 0.0000000714 (days) and central transit time of $T_0=$ 2457599.551478 $\pm$0.000049 (BJD$_{TDB}$).  We find a very good fit with a linear ephemeris (see Fig. \ref{fig:timing}), having a $\chi^2$ value of 16.8 for 21 degrees of freedom (DOF).  We find no obvious signatures of quasi-periodic photometric variabiltiy due to stellar activity, with the TESS data binned by 62 minutes (excluding the transit and eclipse times) giving a standard deviation about the mean of 0.07\% which is only modestly higher than the expected photometric precision of 0.02\%. 

\subsection{White light curve fits}\label{sec:NUVwlc}
\begin{figure*}
	\centering
	\includegraphics[width=\linewidth]{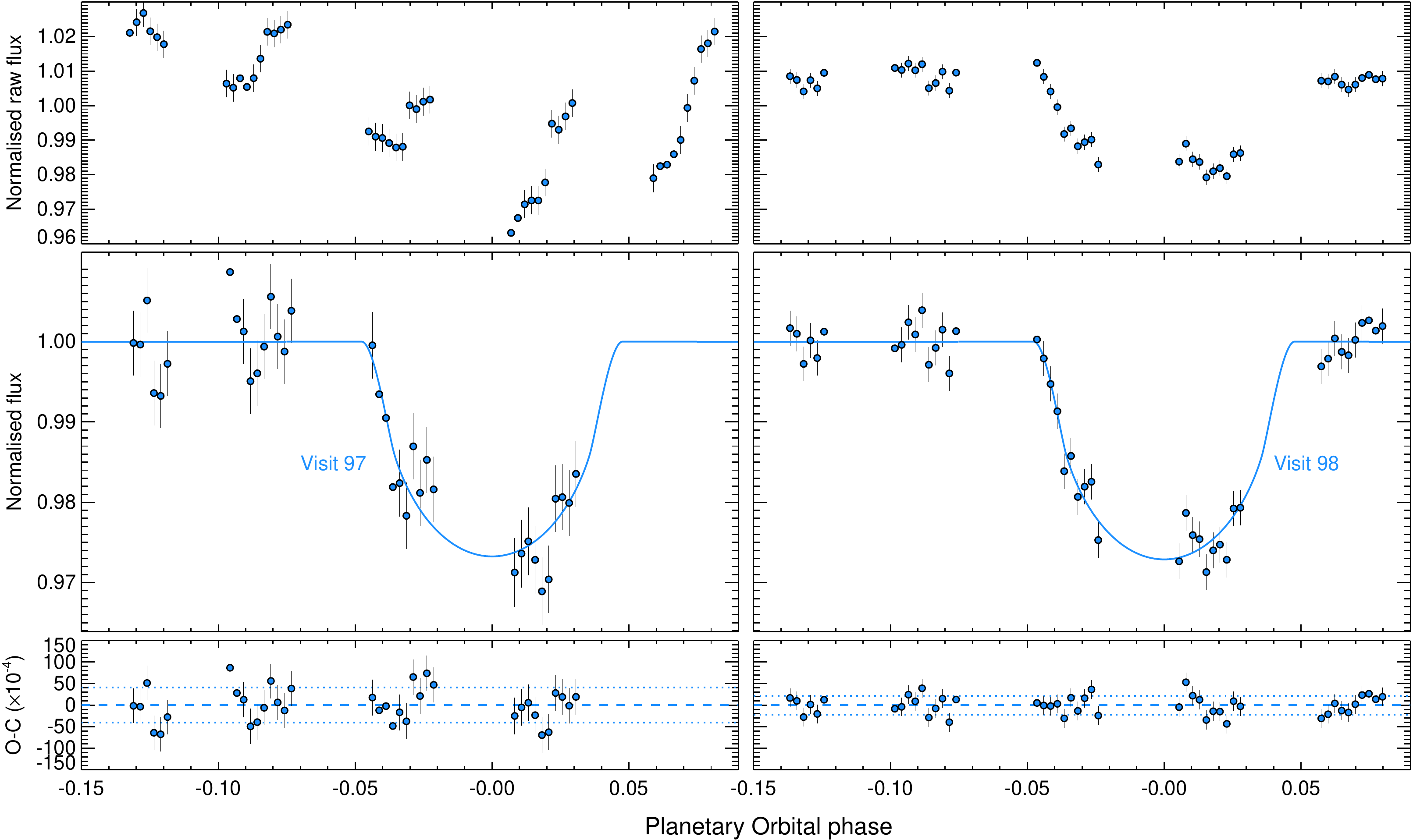}
	\caption{WASP-121b E320M white light curves for visits 97 (left) and 98 (right). Top plot contains the raw flux, the middle panels contain the flux with the fitted instrument systematics model removed, and the bottom panel shows the residuals between the data and the best-fit model.} \label{fig:WLC}
\end{figure*}

\begin{figure*}
	\centering
\vspace{0.25cm}          
	\includegraphics[width=\linewidth]{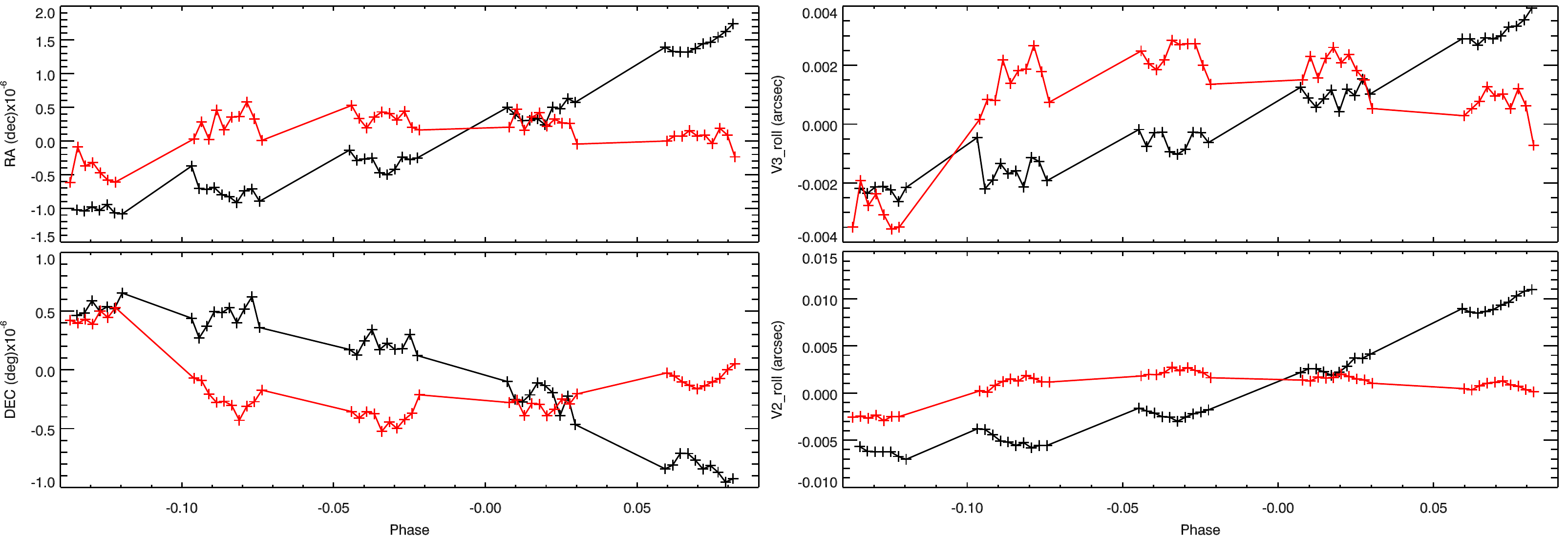}
	\caption{The relative telescope $RA$, $Dec$, $V2_{roll}$, and $V3_{roll}$ for visit 97 (black) and visit 98 (red).} \label{fig:JitterW121}
\end{figure*}

Visits 97 and 98 show dramatically different levels of instrumental systematic trends (see Fig. \ref{fig:WLC}).  In particular, visit 97 shows a $\sim$4\% change in flux in the 5th out-of-transit orbit, while for visit 98 there is no such large trend, and the light curve points are within $\sim$0.4\% of each other.  Upon inspection of the telescope $RA$ and $DEC$ from the jitter files, it is clear that visit 97 suffers from a very large drift during the 5-orbits covering the transit event, while visit 98 has much more stable pointing (see Fig. \ref{fig:JitterW121}).  The large drift likely leads to large slit light losses in visit 97.  Most importantly, the $RA$ and $DEC$ positions of the telescope during the transit in visit 97 are significantly different than the out-of-transit positions, especially the 5th orbit that was shifted by about a quarter of a pixel.  As discussed in \cite{2011MNRAS.411.2199G}, detrending with optical state vectors can only be expected to work if we interpolate the vectors for the in-transit orbit(s), with the out-of-transit orbits, which requires the baseline function to be well represented by the linear model over the range of the decorrelation parameters.

For each visit, we used the AICc and measured $\sigma_{\rm r}$ to determine the optimal optical state vectors to include from the full set without overfitting the data while minimizing the rednoise.  As found in section \ref{sec:jitter}, the different position related vectors from the Jitter files typically contained similar trends.  For visit 98, we found it was optimal to include $\phi_{HST}$ terms up to a second order, as well as the Jitter Detrending vectors $V2\_roll$, $V3\_roll$, and $RA$. We rotated the roll vector pair ($V2\_roll$, $V3\_roll$) which are contained in the engineering jitter files relative to the axis of the square slit ($Vn\_roll$, $Vt\_roll$), which is rotated by about 45 degrees relative to the spacecraft orientation reference vector U3, such that $S(\mathbf{x})$ could be written as,
\begin{multline} 
S(\mathbf{x})=p_1\phi_{t}+p_2\phi_{HST}+p_3\phi_{HST}^2+\\
p_4RA+p_5Vn_{roll}+p_6Vt_{roll}+1.
\label{eq:sys}
\end{multline} 
The fit values of interest, namely $R_{pl}/R_{star}$, did not substantially change for any of the top fitting models (AICc-AICc$_{\mathrm{min}}\lesssim$6).  With nine free parameters ($R_{pl}/R_{star}$, $F_0$, $T_0$, $p_{1..6}$), the fit achieves a signal-to-noise which is 81\% of the theoretical photon noise limit, with no detectable rednoise and a $\chi_{\nu}^2$ of 1.21 for 37 DOF.  We note that
including the jitter corrections and the 2$^{\rm{nd}}$ order breathing polynomial not only provides a significantly better fit than the traditional systematics model, but we were also able to make use of the first orbit in the visit.  The overall photometric performance is similar to that achieved with the STIS CCD using the G430L or G750L, despite the use of a much narrower slit.  With visit 97, we measure the white light curve $R_{pl}/R_{star}=0.1374\pm0.0026$ (see Table \ref{Table:LDTrans}) and a $T_0=$2457854.536411$\pm$0.001073 (BJD$_{TDB}$).
The central transit time agrees very well with the expected ephemeris, occurring -0.2$\pm$1.6 minutes relative to the expected central transit time using the updated ephemeris from Section \ref{sec:tess}.

For visit 97, even the most complex model did not achieve fit residuals as small as visit 98.  
As noted above, the fifth orbit of the visit displays dramatically increased systematic trends compared to the other orbits, with the position of the telescope during the orbit being relatively far from the position of the in-transit orbits (see Fig. \ref{fig:JitterW121}).  
As such, we find the $R_{pl}/R_{star}$ can change dramatically (by about 0.016 $R_{pl}/R_{star}$) between differing systematics models when including the data from the fifth orbit in the light-curve fits.   
Given these factors, we find that it is preferable to drop the last orbit from the analysis when measuring the absolute transit depths in visit 97.  This is similar to many {\it HST} STIS and WFC3 transit studies that have dropped the first orbit as it usually displays larger non-repeatable trends due to changes in the thermal breathing or charge trapping. 
When excluding the fifth orbit, we found the fit $R_{pl}/R_{star}$ values did not change substantially between differing systematics models ($\sim$0.008 $R_{pl}/R_{star}$), and using the AICc, we find $S(\mathbf{x})$=$S(RA$, $Vn_{roll}$, $Vt_{roll}$, $\phi_{t}$) to be optimal.  The best-fit planet radius for visit 97 is found to be $R_{pl}/R_{star}=0.1364\pm0.0110$, which matches the value from visit 98 at well within 1-$\sigma$, and achieves 43\% of the theoretical photon noise limit.


\begin{figure}
\begin{centering}
\includegraphics[width=0.49\textwidth]{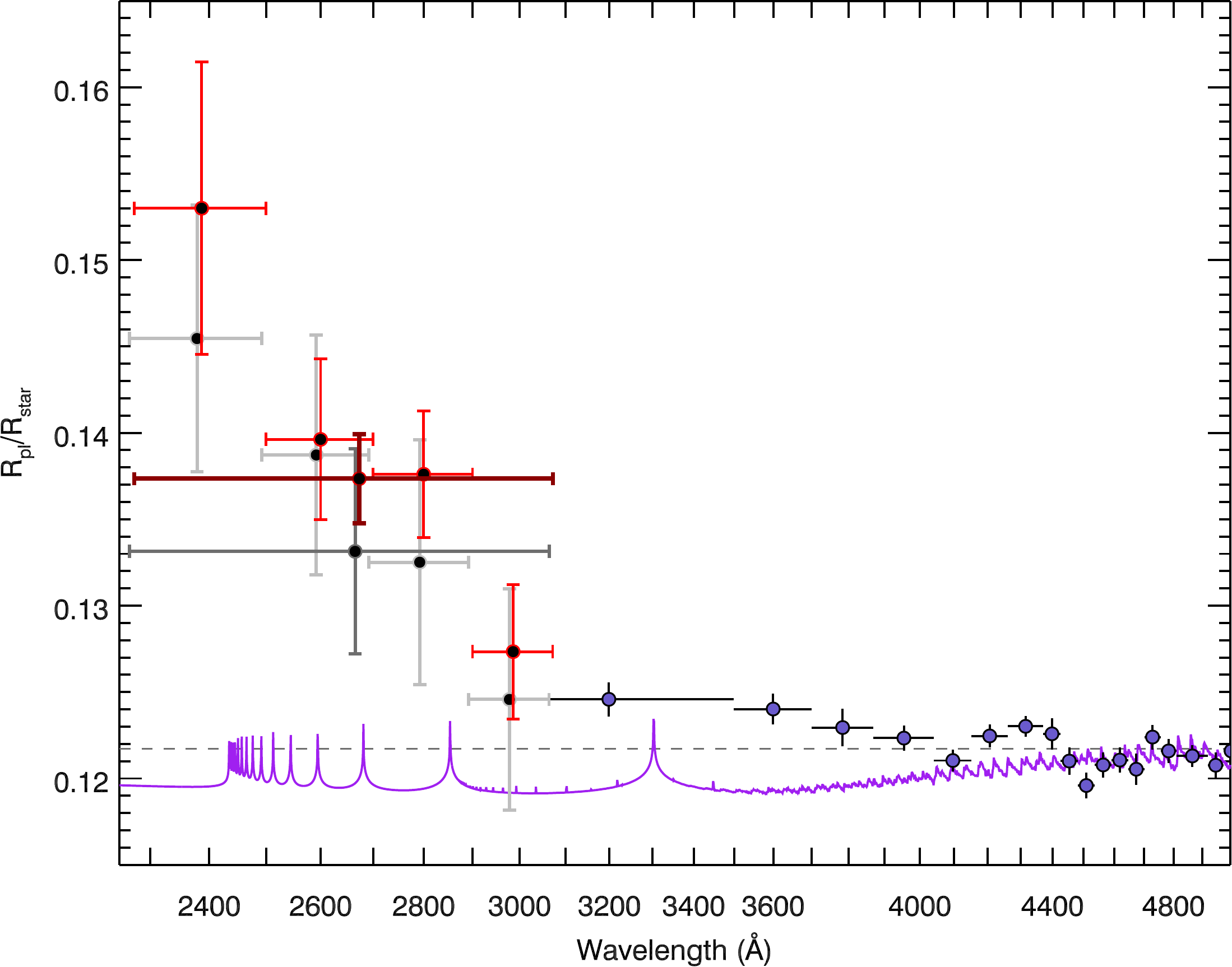}
\par\end{centering}
	\caption{The STIS E230M NUV spectra of WASP-121b compared to the optical transmission spectrum. Plotted (red) is the broadband transmission spectra from visit 98 as well as the white light curve value (dark red).  The broadband spectra NUV spectra from visit 97 is also plotted (grey) along with the white light curve value (dark grey).  The optical spectra are also shown (blue) along with a lower atmospheric model from \cite{2018AJ....156..283E} covering $\sim$mbar pressures (purple) which was fit to the spectra long-ward of 4700 \AA.  The average optical to near-infrared (OIR) value of $R_{pl}(\rm{OIR})/R_{star}=0.1217$ is also shown by the dashed horizontal line. } \label{fig:NUVlr}
\end{figure}

\begin{figure*}
	\centering
\vspace{0.25cm}          
	\includegraphics[width=\linewidth]{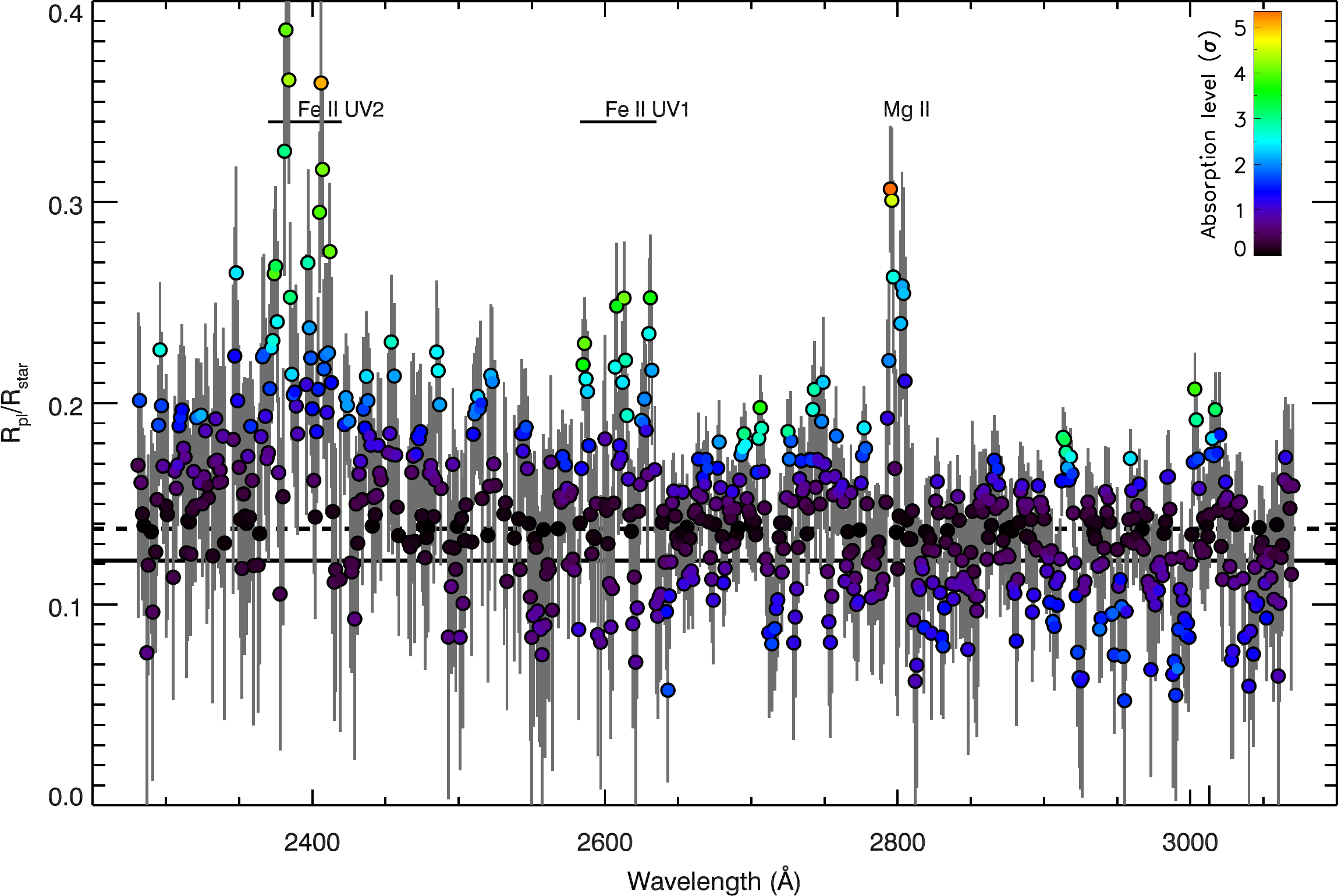} 
	\caption{WASP-121b NUV transmission spectrum scanned in 4 \AA\ passbands, with a 1 \AA\ wavelength shift between each adjacent point.  The data points have been colored with the significance in transit depth above/below that of the NUV broadband value, which is indicated by the dashed line at $R_{pl}(\rm{NUV})/R_{star}=0.1374$.  The average optical to near-infrared (OIR) value of $R_{pl}(\rm{OIR})/R_{star}=0.1217$ is also shown by the solid black line.} \label{fig:scan}
\end{figure*}

\begin{table} 
\caption{WASP-121b broad-band transmission spectral results and non-linear limb darkening
  coefficients for the STIS E230M.}
\label{Table:LDTrans}
\begin{centering}
\begin{tabular}{ccccrrrr}
\hline\hline  
 $\lambda_c$ &$\Delta\lambda$ &  $R_{P}/R_{*}$  &$\sigma_{R_{P}/R_{*} }$ &$c_1$ & $c_2$ & $c_3$ & $c_4$ \\
\AA & \AA\\
\hline  
 2673 &   799   &  0.13735 &   0.00257&0.4011 &-0.2625&1.2818&-0.4674\\
 2387 &   236   &  0.15300 &   0.00845&0.4282 &-0.8777&1.6772&-0.2513\\
 2600 &   200 &  0.13962 &   0.00465&0.4826 &-0.6976&1.9529&-0.7708\\
 2800 &   200   &  0.13760 &   0.00366&0.4090 &-0.2445&1.1797&-0.3940\\
 2986 &    172  &   0.12734 &   0.00389&0.3374 &0.2045 & 0.8312&-0.4348\\
\hline
\end{tabular}
\end{centering}
\end{table}

\subsection{Spectroscopic fits}
When measuring the transmission spectrum, $R_{pl}(\lambda)/R_{star}$, we fixed the system parameters as they are not expected to have a wavelength dependence.  In addition, we fixed the
the limb darkening coefficients to those determined from the stellar model for each spectroscopic passband, using the same method as described in Section \ref{sec:Limbd}.  The light curve fitting methods were otherwise identical to those described in section \ref{sec:lcfits}.   
We fit each spectroscopic light curve with the same functional systematics model as was found to optimally fit the white light curve, fitting the various spectroscopic bins simultaneously for the wavelength-dependent $R_{pl}(\lambda)/R_{star}$ and systematic parameters $p_{1..6}$.

\begin{figure*}
	\centering
	\includegraphics[width=\linewidth]{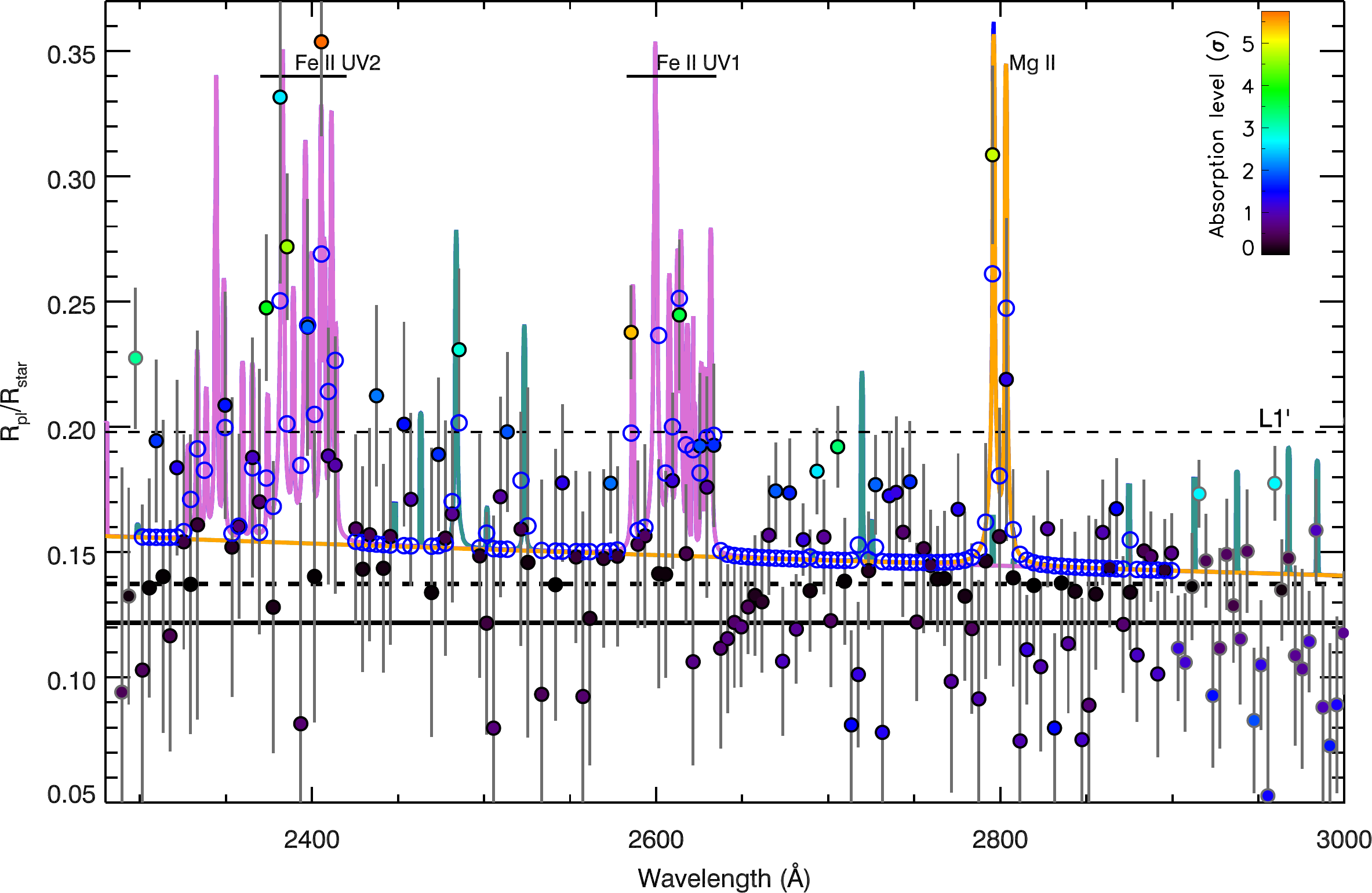} 
	\caption{WASP-121b NUV transmission spectrum in unique 4 \AA\ passbands.  The data points have been colored with the significance in transit depth above that of the NUV spectra, which is indicated by the black dashed line at $R_{pl}(\rm{NUV})/R_{star}=0.1374$.  A model fit that includes Fe\,{\sc i} (green), Fe\,{\sc ii} (purple) and Mg\,{\sc ii} (orange) absorption is shown, with the complete model integrated over the spectral passband shown with (blue) open circles.  The Lagrange point distance L1$'$ is also indicated.\\} \label{fig:waveshift}
\end{figure*}

\begin{figure}
\begin{centering}
\includegraphics[width=0.49\textwidth]{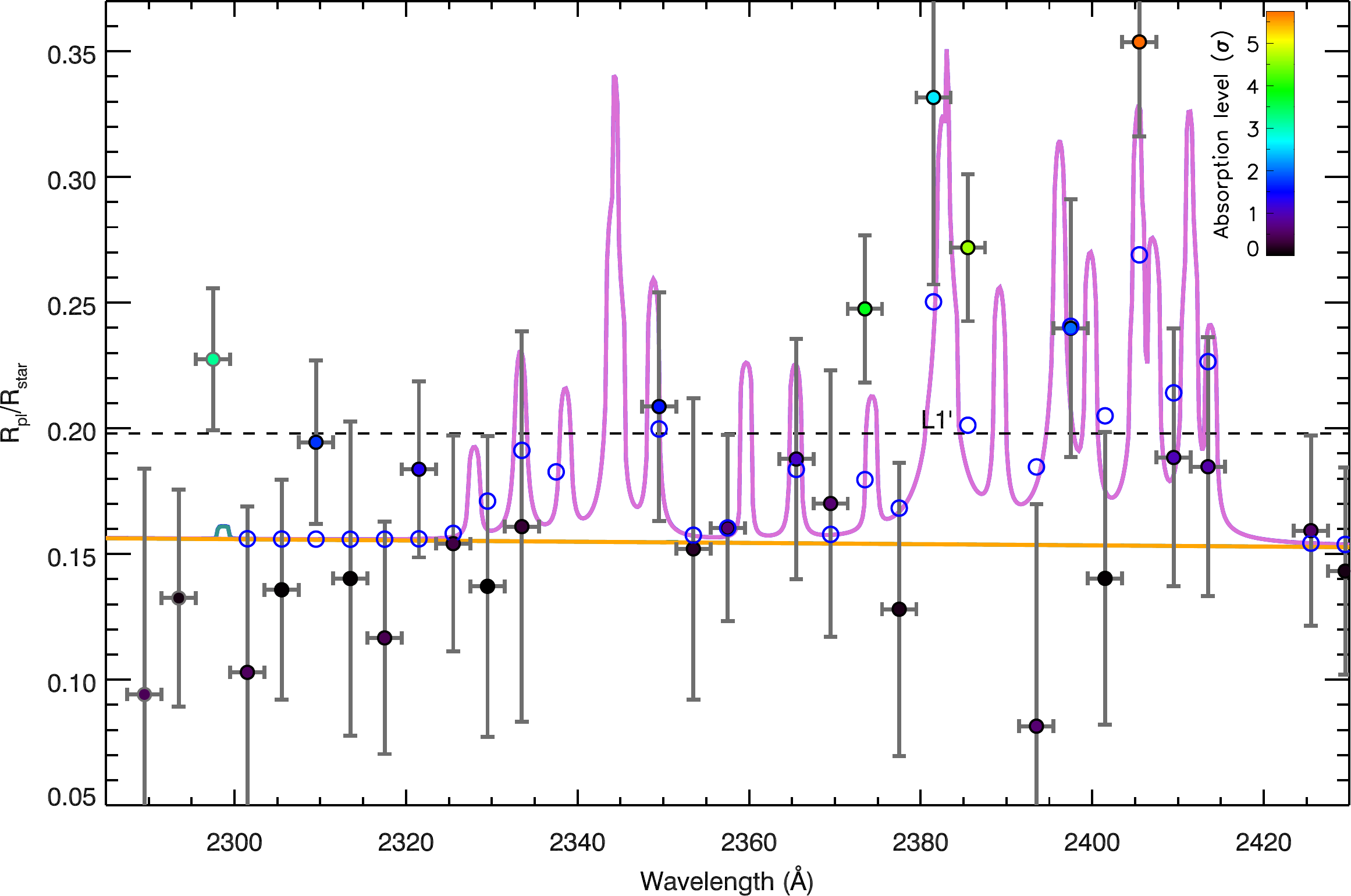}
\par\end{centering}
	\caption{Same as \ref{fig:waveshift} but zoomed in over the Fe\,{\sc ii} UV2 region.  Points have been removed in the spectral regions with strong stellar absorption lines, where the transit itself has not been detected due to low signal-to-noise.} \label{fig:waveshiftA}
\end{figure}

\begin{figure}
\begin{centering}
\includegraphics[width=0.49\textwidth]{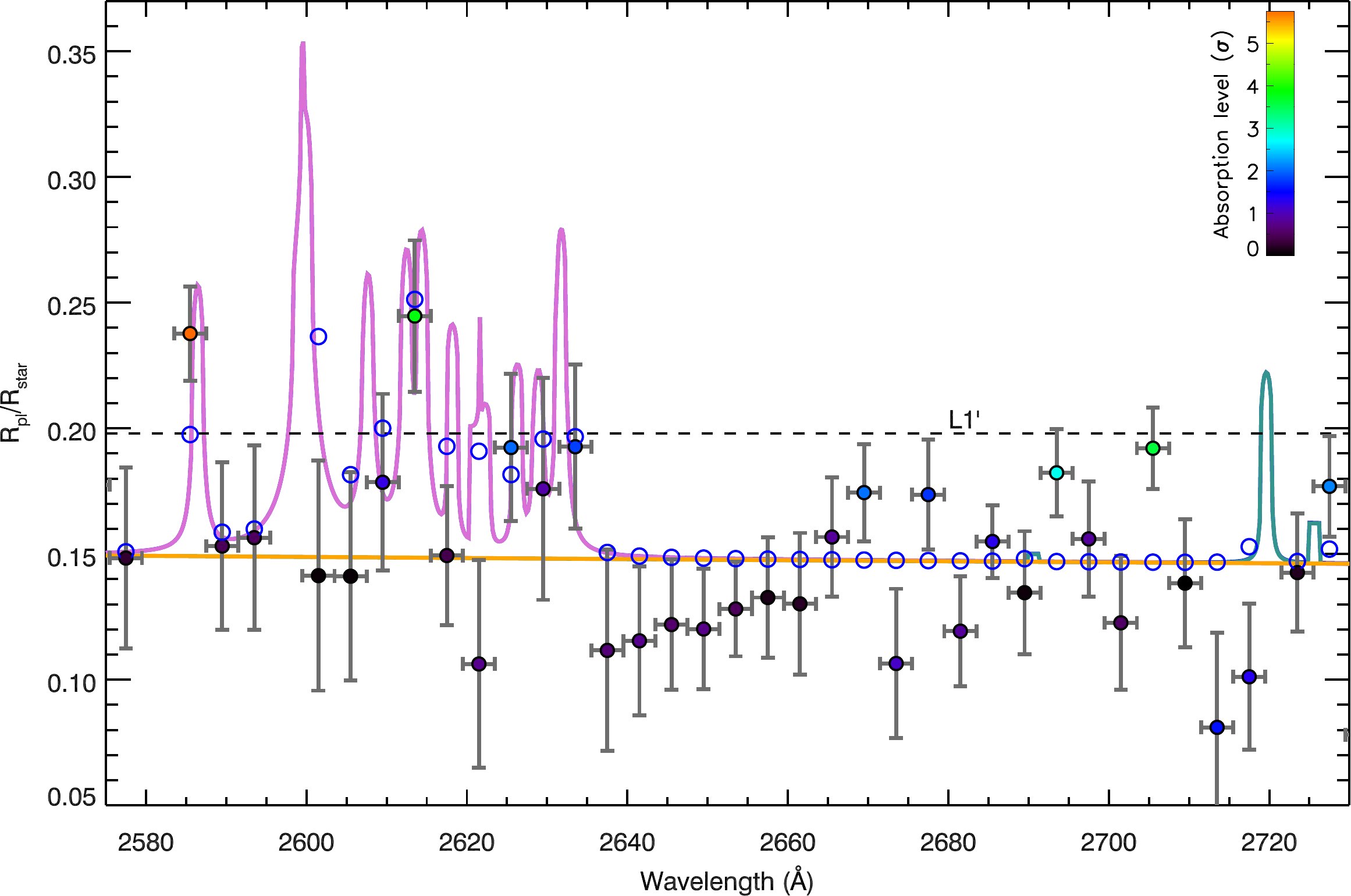} 
\par\end{centering}
	\caption{Same as \ref{fig:waveshift} but zoomed in over the Fe\,{\sc ii} UV1 region. } \label{fig:waveshiftB}
\end{figure}

\begin{figure}
\begin{centering}
\includegraphics[width=0.49\textwidth]{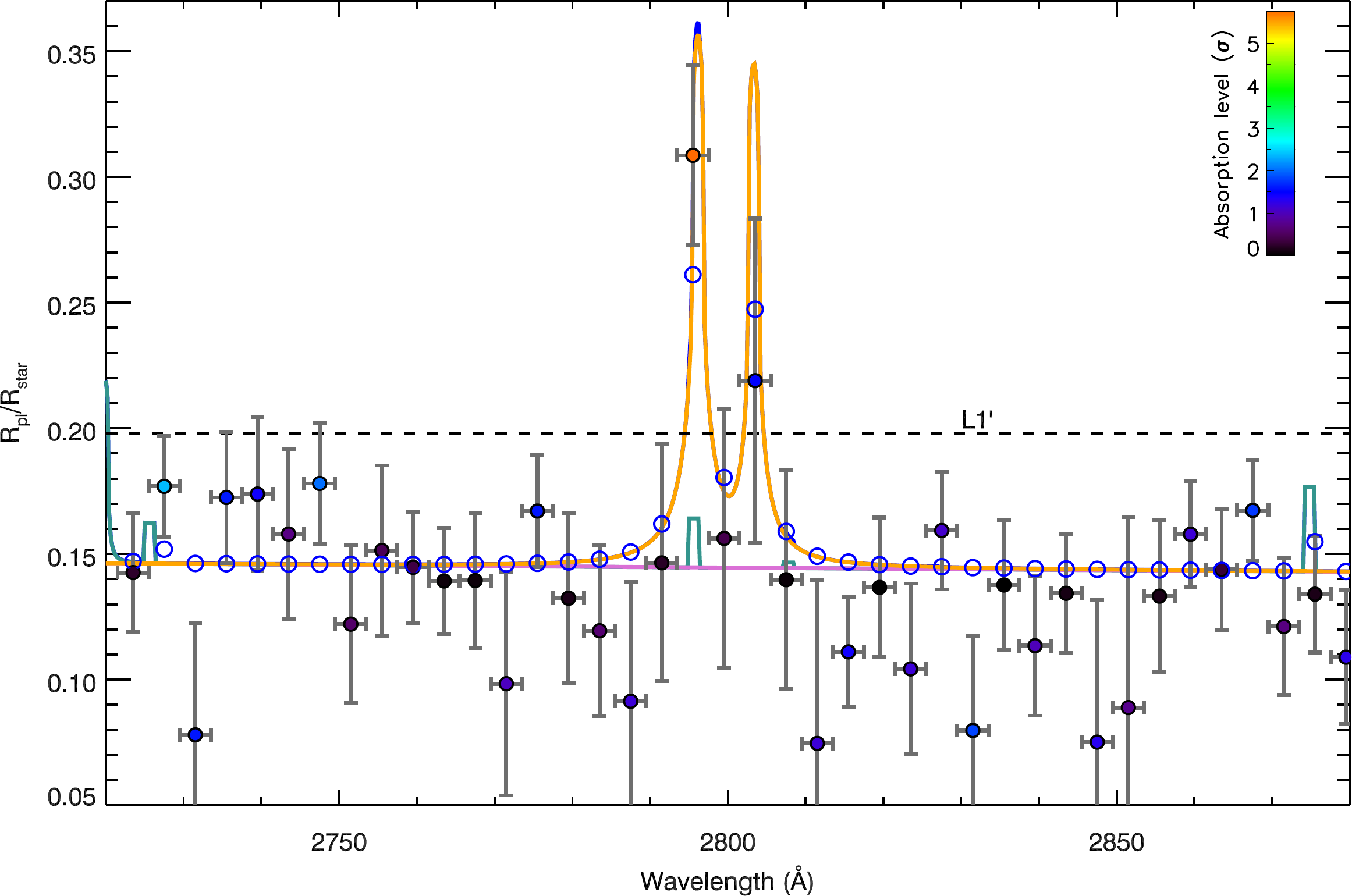}
\par\end{centering}
	\caption{Same as \ref{fig:waveshift} but zoomed in over the Mg\,{\sc ii} region. } \label{fig:waveshiftC}
\end{figure}

Fig. \ref{fig:NUVlr} shows our resulting broadband spectra from visit 98 in $\sim$200 \AA\ bins, as well as the NUV white light curve value.  $R_{pl}(\lambda)/R_{star}$ is observed to sharply rise toward shorter wavelengths.  The overall NUV transmission spectrum shows strong extinction and is substantially higher than the optical and near-IR, reaching altitudes of $\Delta R_{pl}(\lambda)/R_{star}$=0.0157$\pm$0.0026 higher in the atmosphere.  This is more than 18$\times$ the size of the pressure scale height of the lower planetary atmosphere,  $H=kT/\mu g$, that is $H/R_{star}$=0.00084 at a temperature $T$ of 2000 K.

While the overall transit depths of visit 97 are dependent upon the choice of $S(\mathbf{x})$, the steep rise in the NUV broadband spectrum can be seen regardless of including or excluding the fifth orbit from the analysis.
When including the fifth orbit, a systematics model of $S(\mathbf{x})$=$S(\phi_{HST}$, $S_{\lambda}$, $RA$, $DEC$, $Vn_{roll}$) produces a fit radius that is both consistent with visit 98 and independent of excluding the fifth orbit.  The resulting broadband transmission spectra is shown in Fig. \ref{fig:NUVlr}, which matches well with the results of visit 98.

Given the much lower precision and the sizeable systematic uncertainty of including or excluding the fifth orbit in visit 97, we elect to report the transmission spectra at higher resolutions using visit 98 only.
We fit the data to various resolutions and report the transmission spectra at a resolution of $R\sim$650, corresponding to 4 \AA\ bins with results from 5 \AA\ bins also reported in Section \ref{sec:mgfe}.  At these binsizes, each light curve point has a few thousand photons and the $\sim$1.5\% transit depth can typically be detected across the E230M wavelength range.  However, in the spectral regions with strong stellar absorption lines, or at the edges of the spectral orders where the efficiency is low, the transit itself is not always detected due to the lack of flux.  At this resolution, strong atomic transitions can be resolved and probed efficiently and a resolution-linked bias is minimized \citep{2017ApJ...841L...3D}.  Given the high resolution on the E230M, each 4 \AA\ bin still contains $\sim 60$ pixels.  To help probe the central wavelength of the observed features more precisely, we also measured the transmission spectra in multiple 4 \AA\ bin sets, each set shifted by 1 \AA, corresponding to about 100 km/s velocity shifts (see Fig. \ref{fig:scan}).  We removed points where the transit was not detected ($R_{pl}(\lambda)/R_{star}<$0.05) or had large errors ($\sigma_{R_{P}/R_{*} }>$0.09), which predominantly occurred within strong stellar lines or at the order edges.  The R$\sim$650 NUV transmission spectrum can be seen in Figs. \ref{fig:waveshift}, \ref{fig:waveshiftA}, \ref{fig:waveshiftB}, and \ref{fig:waveshiftC}.

\subsection{Mg\,{\sc i}, Mg\,{\sc ii}, Fe\,{\sc i} and Fe\,{\sc ii}}\label{sec:mgfe}
We find no evidence for absorption by Mg\,{\sc i}.  In a 5\AA\ band centered on the ground-state Mg\,{\sc i} line at 2852.965 \AA, we measure a $R_{pl}($Mg\,{\sc i}$)/R_{star}$=0.100$\pm$0.056, which is consistent at the 1-$\sigma$ confidence level with the optical and near-infrared value.

\begin{figure}[ht]
\begin{minipage}[b]{.47\textwidth}
\centering
\includegraphics[width=\linewidth]{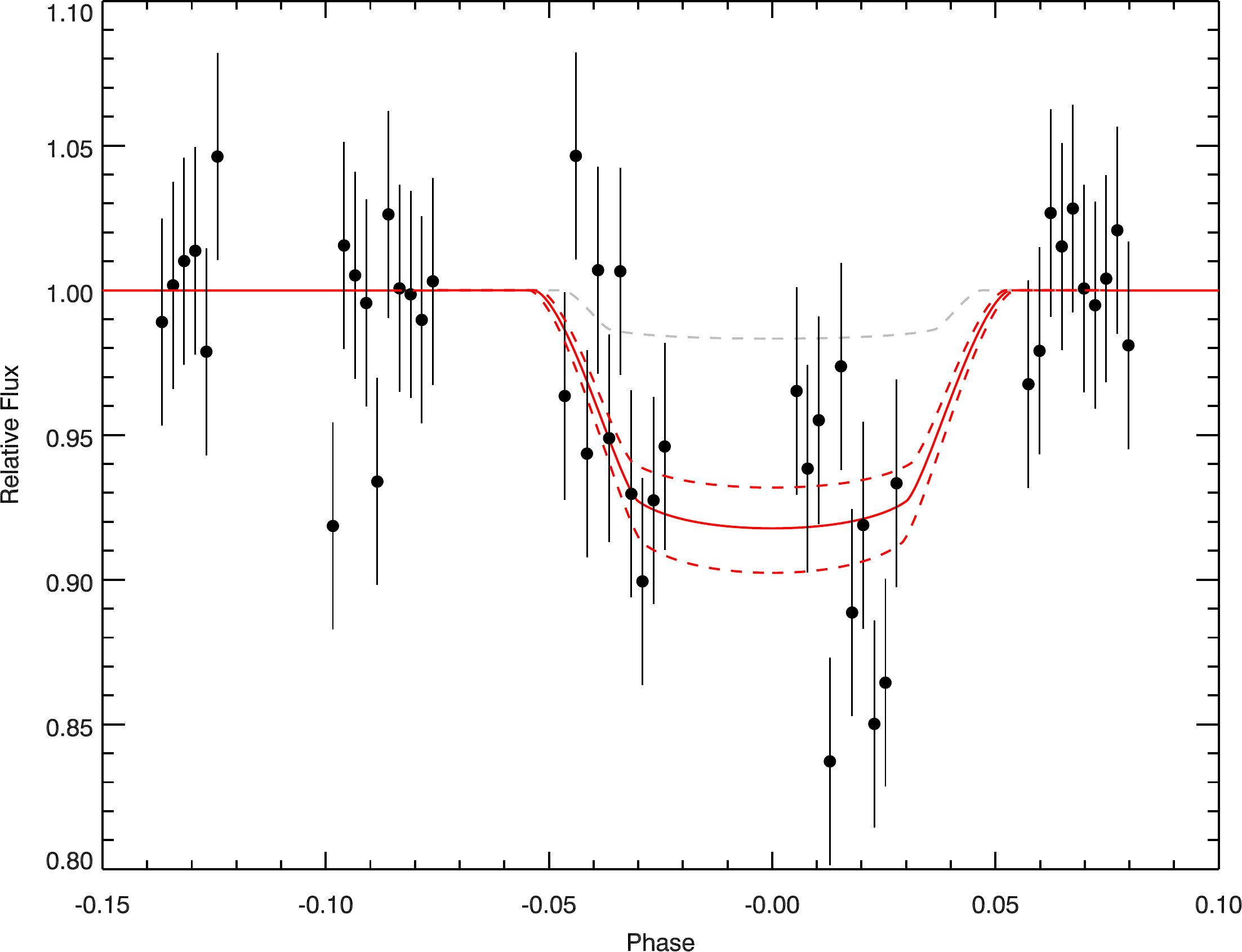}
\end{minipage}
\hfill
\begin{minipage}[b]{.47\textwidth}
\centering
\includegraphics[width=\linewidth]{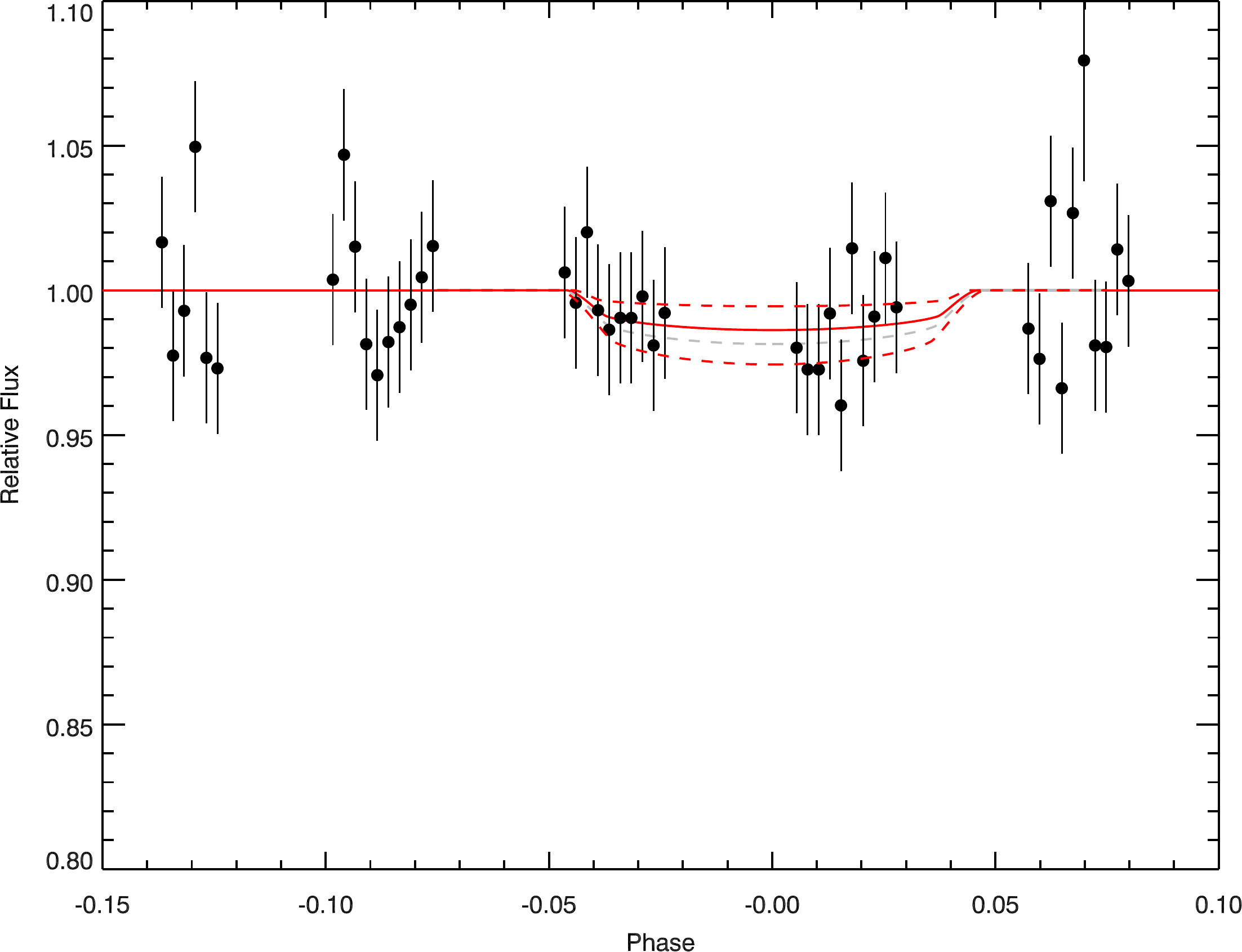}
\vspace*{-0.25\baselineskip}
\caption{Transit light curves for 10 \AA\ wide passbands, which are corrected for systematic errors.  (Top) A light curve composed of two 5\AA\ bins that are centered directly on the Mg II doublet, which optimally covers the two lines. (Bottom) A light-curve composed of two 5 \AA\ bands placed adjacent to either side of the Mg II centered bands, which samples the nearby continuum.  In both plots the average optical-near-IR radius is shown as the grey dashed lines.  The red lines show the best-fit transit depths (solid) and the depths covering the 1-$\sigma$ uncertanties (dashed).}\label{fig:Mgtransits}
\end{minipage}
\end{figure}

We find strong evidence for absorption by Mg\,{\sc ii}.  We detect and resolve absorption by both the k and h features of the Mg\,{\sc ii} ground-state doublet located at 2796.35 and 2803.53 \AA\ respectively (see Fig. \ref{fig:waveshiftC}).  
In 5 \AA\ passbands, the transmission spectra in the Mg\,{\sc ii} doublet is found to have radii of $R_{pl}($Mg\,{\sc ii},k$)/R_{star}=0.284\pm0.037$ and $R_{pl}($Mg\,{\sc ii},h$)/R_{star}=0.242\pm0.0431$, which is substantially larger than the value at optical and near-IR (OIR) wavelengths ($R_{pl}(\rm{OIR})/R_{star}=0.1217$) and larger than the average NUV value of $R_{pl}(\rm{NUV})/R_{star}=0.1374\pm0.0026$.  A 10 \AA\ passband split to cover the Mg\,{\sc ii} doublet simultaneously is found to have a radius value of $R_{pl}($Mg\,{\sc ii},h,k$)/R_{star}=0.271\pm0.024$ (see Fig. \ref{fig:Mgtransits}), which is 5.4-$\sigma$ above the white light curve $R_{pl}(\rm{NUV})/R_{star}$ value.  In the continuum region surrounding the Mg\,{\sc ii} doublet, no other substantial absorption features are observed and most all of the spectral bins in a 100 \AA\ region around the doublet are consistent with the $R_{pl}(\rm{OIR})/R_{star}$ value (see Fig. \ref{fig:waveshiftC}).  

The strong Mg\,{\sc ii} absorption can also be seen in the transit light curves, with Fig. \ref{fig:Mgtransits} showing transit depths of about 8\% within the Mg line, while the surrounding continuum is consistent with the optical and near-IR transit depth of 1.5\%.  The stellar Mg\,{\sc ii} double line cores exhibit narrow emission lines, associated with the stellar corona (see Fig. \ref{fig:3Dmodel}) with both peaks found just red-ward of the line centers at velocities near 20 to 40 km/s.  We performed several checks to ensure that the presence of the emission line did not adversely affect the planetary Mg\,{\sc ii} signal.  Inspecting the photometric time series of just the stellar Mg\,{\sc ii} emission lines, we do not observe any substantial variability differences from that of the surrounding continuum.  In addition, when scanning the transmission spectrum over the Mg\,{\sc ii} region, we find the peak of the transmission spectrum is found 50 to 100 km/s blue-ward of the Mg\,{\sc ii} line cores, rather than red-ward where the emission lines are located.

\restylefloat*{figure}

The transmission spectrum contains a very strong absorption feature at 2381.5 \AA\ ($R_{pl}$(2381.5)$/R_{star}=0.332\pm0.074$), which can be identified with a ground-state resonant transition of  Fe\,{\sc ii}.  Fe\,{\sc ii} transitions occur in a set of multiplets, with notable transitions at 2600, 2382, and 2344 \AA\ for the UV1, UV2, and UV3 multiplets respectively.    For the aforementioned UV1 and UV3 ground-state transitions, the stellar line is strong enough such that almost no flux is measured in the line cores (see Fig. \ref{fig:3Dmodel}), so the transmission spectrum of the planet can not be measured with sufficient precision at those wavelengths.  However, other candidate Fe\,{\sc ii} features are also seen in the stellar spectrum (see Figs. \ref{fig:waveshift}, \ref{fig:waveshiftA}, \ref{fig:waveshiftB}).

For Fe\,{\sc i}, there are strong ground-state transitions in the NUV region at wavelengths of 2484, 2523, 2719, 2913, and 2937 \AA.  While a significant absorption feature does appear near 2484 \AA\ (see Fig. \ref{fig:waveshift}), the other transitions of Fe\,{\sc i} do not obviously appear in the data.

\subsection{1D Transmission Spectra Model}
To help further identify absorption features, we fit a 1D analytic transmission spectral model to the data \citep{2008A&A...481L..83L}, following the procedures as detailed in \cite{2015MNRAS.446.2428S} and  \cite{2018arXiv180407357S}.  Formally, the analytic model is isothermal and assumes a hydrostatic atmosphere with a constant surface gravity with altitude. These assumptions are not expected to be valid over the large altitude ranges probed by the NUV data, especially at very high altitudes.  However, the model is useful for identifying spectral features in the transmission spectra, ruling out absorption by different species, and getting a zeroth-order handle on the atmospheric properties, including the velocity.  A more comprehensive and physically motivated modeling effort will be presented in a future work (P. Lavvas et al. in prep.).

To model the transmission spectra, a continuum slope was included, which was assumed to have cross section opacity with a power law of index $\alpha$,  ($\sigma(\lambda)=\sigma_0(\lambda/\lambda_o)^{-\alpha}$).  We also included the spectral lines of Mg\,{\sc i},  Mg\,{\sc ii}, Fe\,{\sc i}, and Fe\,{\sc ii} with the wavelength-dependent cross-sections calculated using the NIST Atomic Spectra Database \citep{NIST_ASD}.  Voigt profiles were used to broaden the spectral lines.  Given our transmission spectra are at very high altitudes, we did not include effects such as collisional broadening and choose instead to fit for the damping parameter governing the line broadening.  We calculated the scale height assuming a mean molecular weight corresponding to atomic hydrogen and fit the model using an isothermal temperature.  We also assumed that the number of atoms existing in the various atomic levels for a given species followed a Boltzmann distribution with the statistical weights given in the NIST database, though this assumption may not be accurate at very low pressures.  To simplify the calculation, we only included transitions where the lower energy level was beneath a threshold.  High energy levels will in practice not be sufficiently populated to produce significant absorption signatures, and for Fe II we found transitions above 0.43 eV (corresponding to 5000 Kelvin) were not prevalent in the data.  However, the excited states of Fe II up to 0.43 eV were needed, as including only ground-state transitions resulted in a much worse fit to the data.  The presence of these excited states indicates high temperatures, as a significant population of Fe atoms are found above the ground state.

We fit the model to the NUV data with a L-M least-squares algorithm and MCMC \citep{2013PASP..125...83E}.  The fit contained five free parameters: the isothermal temperature, $\alpha$, the Doppler velocity of the planetary atmosphere, the reference planet radius, and the Mg and Fe abundances.  We fit for the data between 2300 and 2900 \AA, finding a good fit with a $\chi^2$ of 130.4 for 126 DOF.  We estimate the detection significance of Fe\,{\sc ii} by excluding its opacity from the model and refitting, finding the $\chi^2$ increases by a $\Delta\chi^2$=61.8, which corresponds to a 7.9-$\sigma$ detection significance.  The velocity of the planetary atmospheric Mg and Fe lines is measured to be $-56_{-63}^{+43}$ km/s.
Given the velocity uncertainties are larger than the spectrograph resolution, the uncertainties can likely be improved using different data reduction methods (e.g. smaller bins or using cross-correlation techniques), which will be presented in a future work (P. Lavvas et al. in prep.).

\section{Discussion}
\subsection{Constraints on the deep interior}
A relatively cold planet interior can lead to the condensing of refractory species deep in the atmosphere, thereby removing the gaseous species from the upper layers of the atmosphere, trapping them within condensate clouds at depth \citep{2003ApJ...594.1011H, 2008ApJ...678.1419F, 2018ApJ...860...18P}, a process that is dependent upon particle settling in the presence of turbulent and molecular diffusion \citep{2009ApJ...699.1487S, 2013Icar..226.1695K}.  Al, Ti, VO, Fe, and Mg are among the first refractory species to condense out of a hot Jupiter atmosphere \citep{2010ApJ...716.1060V, 2017MNRAS.464.4247W}.  As such, the presence or absence of these species can then provide constraints to the temperatures of the deeper layers for hot Jupiters, as the presence of these elements in the gas phase of the upper layers limits the global temperature profile to be hotter than the condensation of these species.  For WASP-121b, the NUV transmission spectra provides evidence for Mg and Fe in the exosphere, while the optical transmission spectrum shows signatures of VO but a lack of TiO \citep{2018AJ....156..283E}. 
The presence of Fe can provide contraints on the interior temperature, $T_{int}$, as the element condenses at the highest temperatures for pressures above about 100 bar (see \citealt{2017MNRAS.464.4247W}).

\begin{figure}
\begin{centering}
\includegraphics[width=0.49\textwidth]{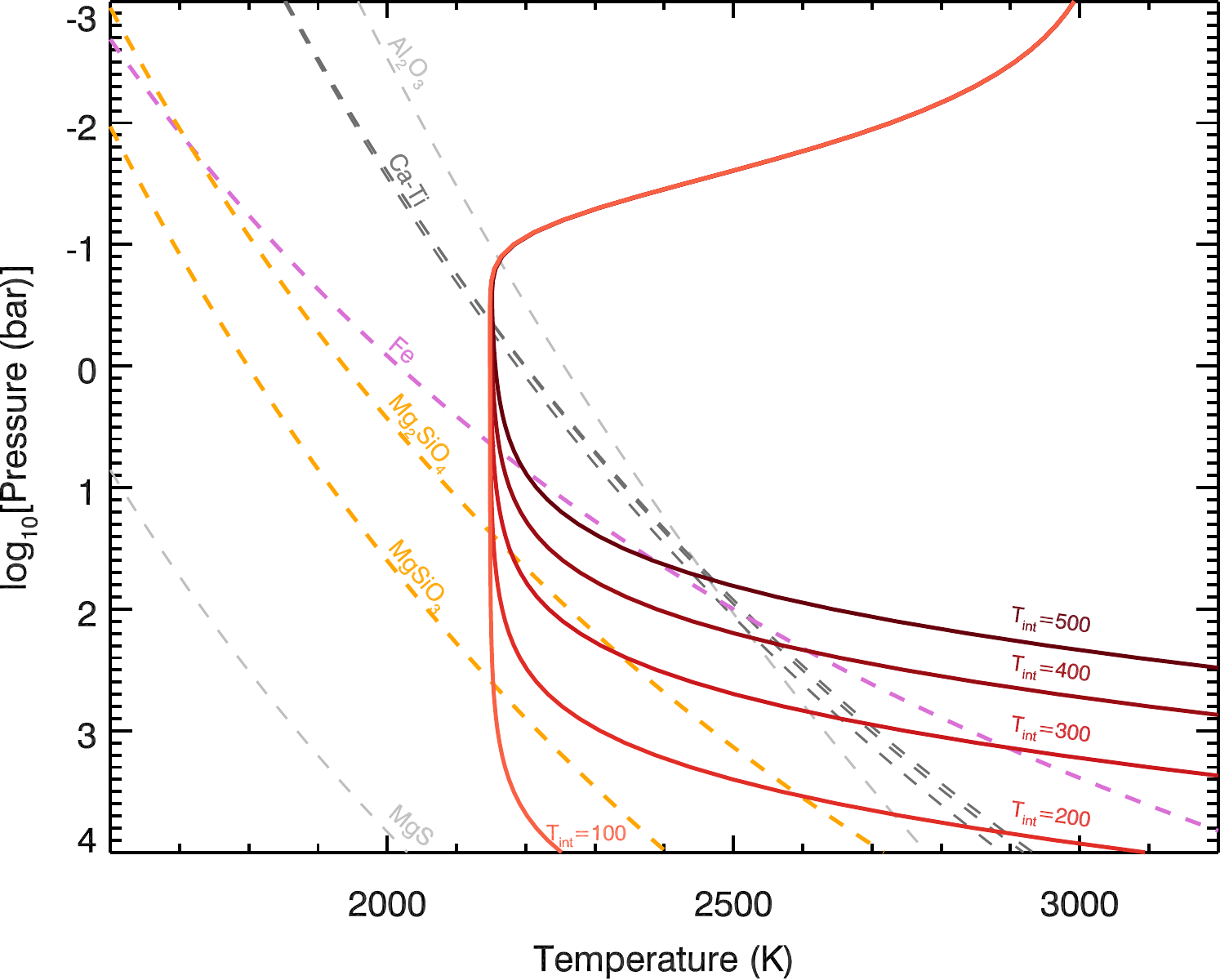} 
\par\end{centering}
	\caption{Temperature-Pressure profiles for WASP-121b (solid lines) compared to condensation curves computed for a metallicity of 10$\times$ solar (dashed lines).  Interior temperatures from 100 to 500 K are shown. } \label{fig:deepinterior}
\end{figure}

 We explore the constraints on the deep interior temperature (pressures $\gtrapprox$ 10 bar) using the best-fit retrieved temperature-pressure (T-P) profile for WASP-121b, derived from fitting the dayside emission spectra of \cite{2017Natur.548...58E}, which is sensitive to approximately bar to mbar pressure levels and has recently been updated to include {\it HST} WFC3 G102 and Spitzer data \citep{2019MNRAS.tmp.1700M}.  The T-P profile was generated using the analytical model of \cite{2010A&A...520A..27G}, which assumes radiative equilibrium, and is parameterized with the Planck mean thermal infrared opacity, $\kappa_{IR}$, the ratio of the optical to infrared opacity, $\gamma$, an irradiation efficiency factor $\beta$, and the interior temperature of the planet, $T_{int}$.  Emission spectra are not directly sensitive to $T_{int}$ \citep{2013ApJ...775..137L}, so retrievals often do not probe the deep interior pressure layers and fix $T_{int}$, with \cite{2017Natur.548...58E} and \cite{2019MNRAS.tmp.1700M}) setting $T_{int}$ to Jupiter-like 100 K values.  However, with both Mg and Fe present in the upper layers of the atmosphere, temperatures of $T_{int}$=100 K are clearly too low, as the T-P profile intersects Fe and Mg condensation curves at high pressures (see Fig. \ref{fig:deepinterior}).   As the condensation curves are metallicity dependent \citep{2010ApJ...716.1060V, 2017MNRAS.464.4247W}, we illustrate the constraints in Fig. \ref{fig:deepinterior} using 10$\times$ solar metallicity, which is close to the best-fit retrieved abundances for WASP-121b in both the transmission and emission spectrum (\citealt{2018AJ....156..283E, 2017Natur.548...58E}, Evans et al. in prep).  Varying $T_{int}$ in steps of 100 K and using the best-fit T-P profile parameters [$\kappa_{IR}$=0.0049 cm$^2$ g$^{-1}$, $\gamma$=4.08, $\beta$=1.03], we find internal temperatures near $T_{int}$=500 K are needed to prevent both Fe and Mg from condensing deep in the atmosphere, but also allows TiO to condense at pressures well below the photosphere. 
As discussed in \cite{2008ApJ...678.1419F}, for adiabatic interiors higher values of $T_{int}$ lead to warmer interiors and larger planet radii.  Given WASP-121b is one of the largest exoplanets found ($R_{pl}=1.865R_{\rm Jup}$,  \citealt{2016MNRAS.458.4025D}), a high $T_{int}$ would be expected.  While there are computational limitations, we note that self-consistent 3D modeling efforts that include both the deep interior and the exosphere in comparison to transmission, emission and phase curve data can provide better constraints on the irradiation efficiency, atmospheric abundances, atmospheric mixing, and global T-P profiles, which in turn would provide the best condensation constraints on $T_{int}$.

\subsection{Roche lobe geometry and exospheric constraints}
The ionized Mg and Fe lines are seen up to extremely high altitudes, corresponding to high planetary radii of $R_{pl}/R_{star}\sim$0.3.  We compare the altitudes of this absorption material to the theoretical distances of the L1 Lagrange point, $D_{\rm L1}$, and the L1$'$ Lagrange point, $D_{\rm L1'}$, which is the distance between the planet's center and the closest point of the equipotential surface including L1, which is the Roche lobe.  Using the formalism in \cite{2003ApJ...588..509G} and using a mass ratio of $M_{pl}/(M_{pl}+M_{star})=8.3\times10^{-4}$ as well as $a/R_{star}=3.76$, we calculate 
$D_{\rm L1}/R_{star}$=0.24 (or 1.96$R_{pl}$) and the L1$'$  Lagrange size relative to that of the star to be $D_{\rm L1'}/R_{star}$=0.209 (or 1.72$R_{pl}$).
In a transit configuration, the observed limit of the Roche lobe is perpendicular to the planet-star direction and the Roche lobe extends to about 2/3 of the extension to the L1 Lagrange point \citep{2008ApJ...676L..57V}.
For WASP-121b, we numerically calculated the equivalent radius\footnote{equivalent meaning the radius of a disk with the same area} of a transiting Roche lobe, R$_{\rm eqRL}$, finding $R_{\rm eqRL}/R_{star}$ = 0.158 (or 1.3 $R_{pl}$).

The core of the Mg\,{\sc ii} k line reaches up to $R_{pl}/R_{star}$= 0.309$\pm$0.036 (or 2.52$\pm$0.29$R_{pl}$), which is 
in excess of $R_{\rm eqRL}/R_{star}$ at 4-$\sigma$ confidence.  In addition, the ground-state Fe\,{\sc ii} resonance line at 2382\AA\ reaches up to $R_{pl}/R_{star}$ =0.331$\pm$0.074 (or 2.7$\pm$0.6$R_{pl}$), which is in excess of $R_{\rm eqRL}/R_{star}$ at 2.7-$\sigma$ confidence.  In a transit configuration, these absorption features indicate that both Mg\,{\sc ii} and Fe\,{\sc ii} reach altitudes such that they are no longer gravitationally bound to the planet.  While large hydrogen tails extending well beyond the Roche lobe have been seen at Lyman-$\alpha$ \citep{2015Natur.522..459E, 2018A&A...620A.147B}, elements heavier than hydrogen in exoplanets have not previously been found at distances in excess of the Roche lobe.  

The existence of heavy atmospheric material reaching beyond the Roche lobe agrees with a geometric blow-off scenario \citep{2004A&A...418L...1L}, where the Roche lobe is in close spatial proximity to the planet and the exobase can extend beyond it letting atmospheric gas freely stream away from the planet.  WASP-121b is an ideal planet to potentially observe this phenomenon, as the planet is on the verge of tidal disruption \citep{2016MNRAS.458.4025D}.  At the terminator, the planet fills a significant portion of it's Roche lobe, with the optical-to-infrared radii reaching $R_{pl}(\rm{OIR})/R_{\rm eqRL}$=77\%.  The NUV reaches $R_{pl}(\rm{NUV})/R_{\rm eqRL}$=87$\pm$2\%, indicating the NUV continuum contains opaque material nearly up to the transit-projected Roche lobe, while the cores of the Mg\,{\sc ii} and Fe\,{\sc ii} exceed the Roche lobe.  A similar process is likely happening in comparably hot and tidally distorted planets such as WASP-12b, which also shows evidence of Mg\,{\sc ii} and Fe\,{\sc ii} absorption \citep{2010ApJ...714L.222F, 2012ApJ...760...79H}.   However WASP-121b has a more favorable transmission spectral signal and orbits a significantly brighter star than WASP-12A.  These properties allow the WASP-121b NUV data to be fit at relatively high-resolution with a full limb-darkened instrument-systematic corrected transit model, which allows the transmission spectral line profiles to be well resolved and absolute transit depths preserved.   Our transit light curves cover ingress, though no early ingress is observed as the NUV transit time is in excellent agreement with the expected ephemeris (see Section \ref{sec:tess}).  Like WASP-12b \citep{2015ApJ...803....9N, 2016MNRAS.459..789T}, there is no evidence in WASP-121b for a bow-shock or material overflowing the L1 Lagrange point \citep{2010ApJ...721..923L, 2010ApJ...722L.168V, 2011MNRAS.416L..41L, 2013ApJ...764...19B}. 

As ionized species, the Mg\,{\sc ii} and Fe\,{\sc ii} atoms are potentially sensitive to the planet's magnetic field, which if strong enough could lead to magnetically controlled outflows \citep{ 2011ApJ...730...27A}.  As such, departures in the transit light curves from spherical-symmetry and the velocity profile of the ionized lines could give insights into the nature of the outflow.  Our data lacks complete phase coverage, so only has a limited ability to constrain a non-spherical absorption profile and we cannot constrain a possible post-transit cometary-like evaporation tail.  However, we note that all of our transit light curve fits were well fit to near the photon-noise limit assuming the planet was a sphere.  In addition, our model fit found velocities of $-56_{-63}^{+43}$ km/s, which does not have sufficient precision to confidently distinguish between outflowing blue-shifted gas or zero velocity gas.

\section{Conclusion}
We have presented {\it HST} NUV transit observations of WASP-121b and have introduced a new detrending method, which we find helps improve the photometric performance of time-series {\it HST} STIS data.  WASP-121b is one of the few exoplanets with a complete NUV-OIR transmission spectrum that can be compared on an absolute transit depth scale from NUV wavelengths into the near-IR. The whole NUV wavelength region shows significant absorption, with transit depths well in excess of the optical and near-IR levels and a NUV continuum that rises dramatically blueward of about 3000 \AA.  We find spectral characteristics that have not previously been observed, with strong features of ionized Mg and Fe lines that extend well above the Roche lobe observed, with multiple resolved spectral lines detected for both species.  While not gravitationally bound, it is unclear whether these ionized species are magnetically confined to the planet, though better signal-to-noise spectra would improve velocity measurements, and complete phase coverage searching for non-spherical symmetries could provide further insight.  The presence of gas beyond the Roche lobe is evidence this tidally distorted planet is undergoing hydrodynamic outflow with a geometric blow-off, where the exobase extends to or exceeds the Roche lobe, and elements heavier than hydrogen are free to escape.  
Spectrally resolved Mg\,{\sc ii} and Fe\,{\sc ii} features on an absolute transit depth scale will allow detailed simulations of the local physical conditions of the upper atmosphere, where it may be possible to deduce the altitude dependent density, temperature, and abundances of the upper atmosphere.

\section*{ACKNOWLEDGEMENTS}
\acknowledgments
This work is based on observations made with the NASA/ESA Hubble Space Telescope that were obtained at the Space Telescope Science Institute, which is operated by the Association of Universities for Research in Astronomy, Inc.
Support for this work was provided by NASA through grants under the HST-GO-14767 program from the STScI
A.G.M. acknowledges the support of the DFG priority program SPP 1992 "Exploring the Diversity of Extrasolar Planets (GA 2557/1-1).
P.L. also acknowledges support by the Programme National de Plan\'etologie (PNP) of CNRS/INSU, co-funded by CNES.
L.B.-J. and P.L. acknowledge support from CNES (France) under project PACES.
J.S.-F. acknowledges support from the Spanish MINECO grant AYA2016-79425-C3-2-P.
This project has been carried out in part in the frame of the National Centre for Competence in Research PlanetS supported by the Swiss National Science Foundation (SNSF), and has received funding from the European Research Council (ERC) under the European Union's Horizon 2020 research and innovation programme (project Four Aces; grant agreement No 724427).
The authors would like to thank the staff at STScI for their extra efforts with these datasets which required special handling and scheduling.  We also would like to thank the anonymous reviewer for their constructive comments and suggestions. 

\vspace{5mm}
\facilities{HST (STIS)}
\facilities{TESS}

\bibliographystyle{yahapj} 
\bibliography{W121_NUV_Mg_Fe_Jitter_Detrending} 


\appendix

\section{Benchmarking Jitter Detrending with WASP-12b STIS CCD Eclipse data}\label{sec:W12appendix}
\begin{figure*}
	\centering
	\includegraphics[width=\linewidth]{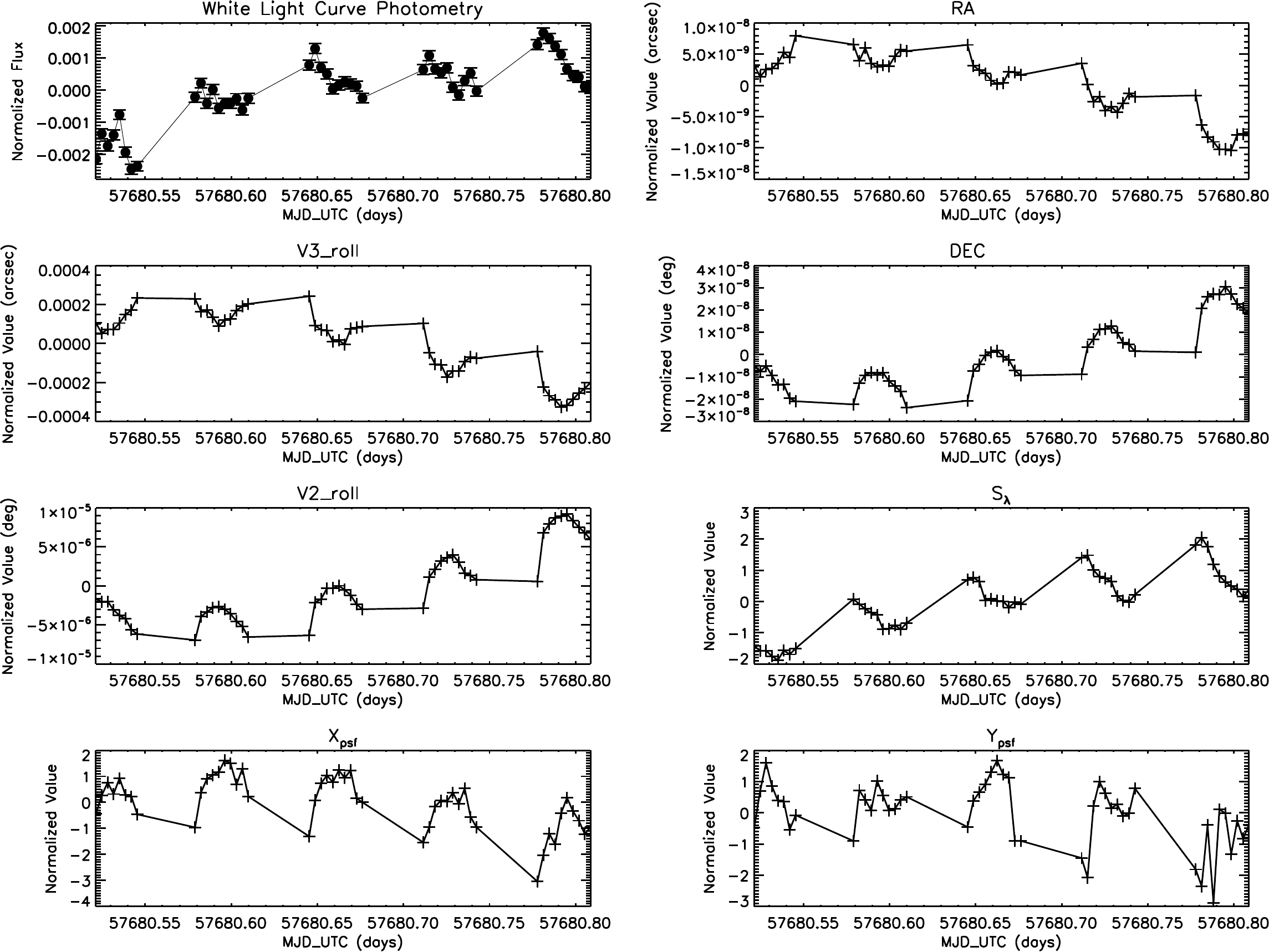}
	\caption{White light curve of WASP-12b and selected jitter engineering measurements ($V3_{roll}$, $V3_{roll}$, $RA$, and $DEC$) for visit 3.  Several traditional detrending variables ($X_{psf}$, $Y_{psf}$ and  $S_{\lambda}$) are shown as well.}\label{fig:jit}
\end{figure*}

We benchmarked the use of jitter detrending on the STIS G430L eclipse data of WASP-12b to verify the jitter engineering data could be used to detrend and fit STIS transit light curve data and to compare the performance with the latest methods.  
A subset of the optical state jitter vectors can be seen in Figure \ref{fig:jit} for the 2017 June 15 STIS data.  Results from these data have been published in \cite{2017ApJ...847L...2B}, who used a Gaussian process with the traditional model optical state vectors to model the light curve systematics.  In our analysis, the data reduction steps were identical to \cite{2017ApJ...847L...2B}; however, we included the first {\it HST} orbit in our analysis and discarded the first exposure in each orbit.

We modelled $N$ measured fluxes over time, $f(t)$, as a combination of an eclipse model, $T(t, \mathbf{\delta})$
(which depends upon the eclipse parameter $\mathbf{\delta}$), the total
baseline flux detected from the star, $F_0$, and a parameterised systematics error
model $S(\mathbf{x})$ giving,
\begin{equation} 
f(t)=T(t, \theta)\times F_0 \times S(\mathbf{x}).
\end{equation} 
With no ingress or egress measurements in the light curves, as done by \cite{2013ApJ...772L..16E} and \cite{2017ApJ...847L...2B}, we used a boxcar function to describe the eclipse signal, with
\begin{equation}
\begin{aligned}
	&T_i = (1 - \delta B_i)\\
	&B_i = 
		\begin{cases} 
    		0 & i \in \text{out of eclipse} \\ 1 & i \in \text{in eclipse}\,,
		\end{cases}
\end{aligned}
\end{equation}%
where $\delta$ is the fractional flux change during the eclipse for exposure measurements $i = 1$, ..., $N$.
For our systematics error model, we included the traditional optical state vectors ($\phi_{HST}$, $\phi_{HST}^2$, $\phi_{HST}^3$, $\phi_{HST}^4$, $X_{psf}$, $Y_{psf}$ and  $S_{\lambda}$) as well as the jitter vectors $V2\_roll$, $V3\_roll$, $RA$, $DEC$, $Lat$, $Long$ such that the most complex model contained thirteen total terms used to describe $S(\mathbf{x})$.  We included only linear terms for all of the optical state vectors aside from $\phi_{HST}$, which was fit up to fourth order.  Higher order terms in the rest the parameters were explored but found not to improve the fits.  
With a $f(t)$ containing only linear parametrised terms, including the eclipse model, we use a multiple linear regression fit to find the best-fit parameters and 1-$\sigma$ uncertainly estimates \citep{2003drea.book.....B}.  For the systematics error models, we fit the light-curves using all combinations of the thirteen optical state vectors terms for $S(\mathbf{x})$ giving $2^{13}=8192$ total fits to the data for each {\it HST} visit.  By fitting all combinations, our model includes not only the traditional decorrelation parameters but models with jitter vectors as well.  Rather than selecting only the best-fitting light curve, we used the marginalization method as described in \cite{2014MNRAS.445.3401G} to incorporate the uncertainty in the choice of the systematics model into the eclipse depth measurement.  This method has been shown to be effective for WFC3 data \citep{2016ApJ...819...10W} and has also been used for STIS data \citep{2018AJ....155...66L}.  For all 8192 systematics models, we approximate the evidence, $E_q$, using the second-order Akaike Information Criterion (AICc),
\begin{equation}
\begin{aligned}
 E_q\approx-\frac{1}{2}\mathrm{AICc}\\
\mathrm{AICc}=\mathrm{AIC}+\frac{2k^2+2k}{n-k-1}
\end{aligned}
\end{equation}
where $N$ denotes the number of measurements in the light curve and $k$ denotes the number of fit parameters.
The AICc applies a correction for small sample sizes to the Akaike Information Criterion (AIC), and so helps to address potential overfitting problems in the case when $N$ is small as can be the case in this data.
Photon noise error bars were initially assumed when fitting the light curves, and the measured parameters were then rescaled based on the standard deviation of the light curve residuals.  

\begin{figure}
	\centering
	\includegraphics[width=\linewidth]{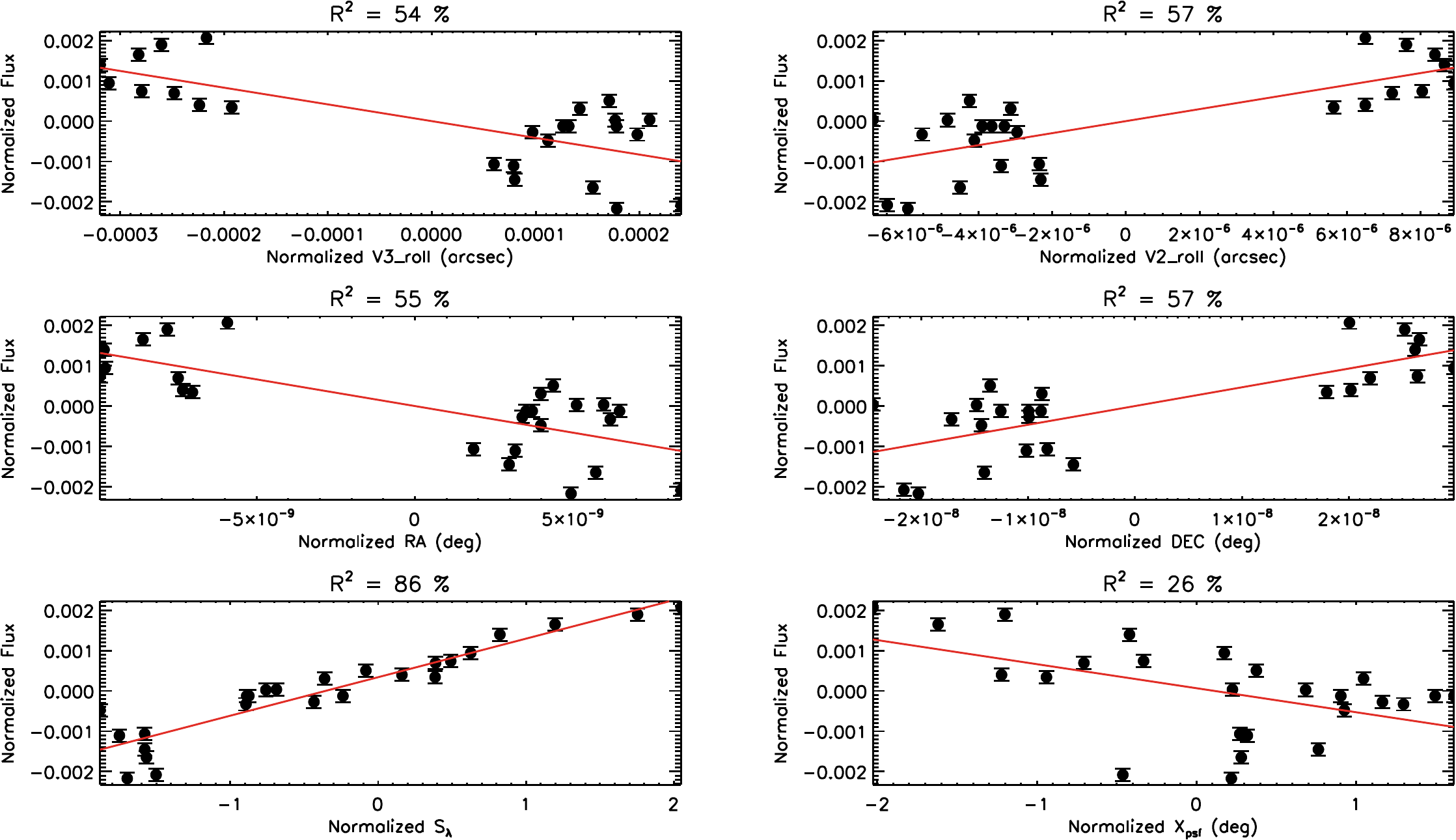}
	\caption{Correlations between the WASP-12b out-of-eclipse white light curve data and several optical-state jitter vectors.  A linear trend (red) and $R^2$ correlation values are also shown.}\label{fig:correlate}
\end{figure}

The vector pairs ($V2\_roll$, $V3\_roll$) and ($RA$, $DEC$) often exhibit very similar trends.  In practice, including multiple vectors with the same (if not identical) trends could bias the model-averaged results when marginalising, as it can result in over-counting the common trends contained in the vectors, and detrending with multiple similar optical vectors will introduce large fitting degeneracies.  Our main objective here was to allow the fitting to use any combination of optical state vectors to directly test the performance of the Jitter optical state vectors against the traditional vectors, so we chose to include all vectors in our study rather than select unique vectors.  However, to help mitigate the aforementioned shortcomings, we used Principle Component Analysis (PCA) to convert the vector pairs into a set of orthogonal vectors, which were then subsequently used to fit the light curves.  This additional PCA procedure helped to reduce fitting degeneracies between highly correlated jitter-vector pairs and also helped effectively reduce the number of jitter vectors that were included in the our best-fitting models.

\begin{figure}[h]
	\begin{minipage}[t]{8.5cm}
		\centering
		\includegraphics[scale=0.4]{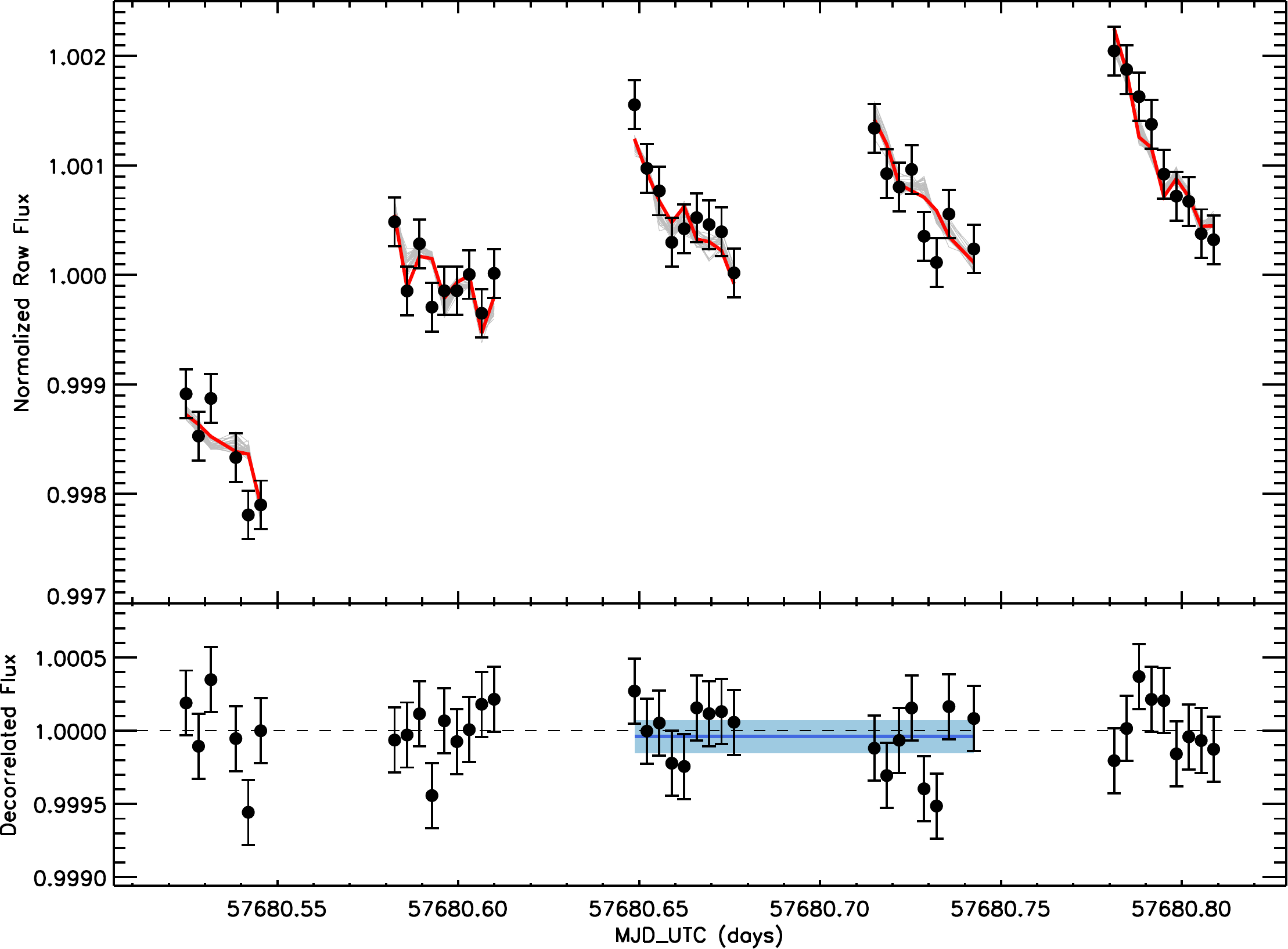}
		\caption{(Top) Raw light curve time series flux of WASP-12b.  The top-fitting model is shown in red, and models with weights larger than 0.5\% are also shown in grey.
(Middle) Detrended light curve, dark blue line indicates the top-fitting model eclipse depth and the 1-$\sigma$ uncertainty is indicated by the light blue bar.}\label{fig:VisitW12a}
	\end{minipage}
	\hspace{0.5cm}
	\begin{minipage}[t]{8.5cm}
		\centering
		\includegraphics[scale=0.4]{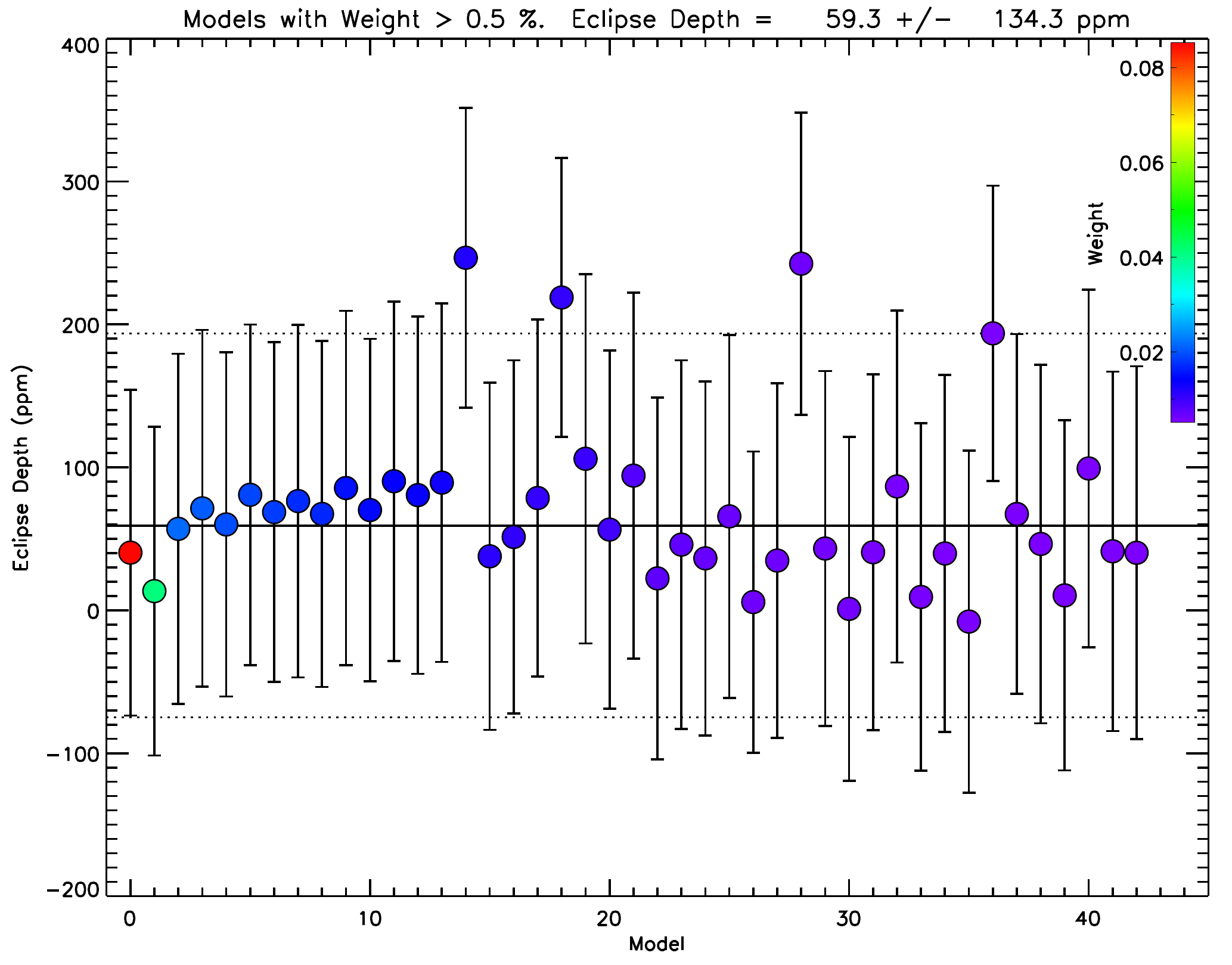}
		\caption{Marginalized eclipse depth results showing models with weights over 0.5\%.}\label{fig:VisitW12b}
	\end{minipage}
 \end{figure}

Our best-fit to the white light curve data achieved precision of 70\% of the theoretical photon noise limit with standard deviations of 218 parts per million (see Figs. \ref{fig:correlate}, \ref{fig:VisitW12a} and \ref{fig:VisitW12b}).  The best-fitting systematics models included the Jitter Detrending optical state vectors $V2\_roll$, $V3\_roll$, $DEC$, $Lat$, and $Long$.  
For the broadband eclipse depth, we measure a model-marginalized eclipse depth of $\delta=59\pm134$, which is compatible within the errors to \cite{2017ApJ...847L...2B} who reported $\delta=-53\pm74$.  As noted by \cite{2017ApJ...847L...2B},  the measured eclipse depth should be strictly positive, though in practice it may be negative due to either random or systematic noise.   With Jitter Detrending, the first orbit was successfully recovered for use in the analysis, and we find the measured eclipse depth is positive.  The eclipse depth translates to a measured geometric albedo of ($A_g=3.9\pm8.8$\%), which is very low and in agreement with the findings of \cite{2017ApJ...847L...2B}.
Thus, with this test and the results of \ref{sec:trends}, we conclude the Jitter products of the EDPS generally contain sufficiently accurate information to help decorrelate precision time-series spectrophotometry.

\end{document}